# Taurus: A Data Plane Architecture for Per-Packet ML

Tushar Swamy
*Stanford University*

Alexander Rucker
*Stanford University*

Muhammad Shahbaz
*Purdue University*

Ishan Gaur
*Stanford University*

Kunle Olukotun
*Stanford University*

## ABSTRACT

Emerging applications—cloud computing, the internet of things, and augmented/virtual reality—demand responsive, secure, and scalable datacenter networks. These networks currently implement simple, per-packet, data-plane heuristics (e.g., ECMP and sketches) under a slow, millisecond-latency control plane that runs data-driven performance and security policies. However, to meet applications' service-level objectives (SLOs) in a modern data center, networks must bridge the gap between line-rate, per-packet execution and complex decision making.

In this work, we present the design and implementation of *Taurus*, a data plane for line-rate inference. Taurus adds custom hardware based on a flexible, parallel-patterns (MapReduce) abstraction to programmable network devices, such as switches and NICs; this new hardware uses pipelined SIMD parallelism to enable per-packet MapReduce operations (e.g., inference). Our evaluation of a Taurus switch ASIC—supporting several real-world models—shows that Taurus operates orders of magnitude faster than a server-based control plane while increasing area by 3.8% and latency for line-rate ML models by up to 221 ns. Furthermore, our Taurus FPGA prototype achieves full model accuracy and detects two orders of magnitude more events than a state-of-the-art control-plane anomaly-detection system.

## 1 INTRODUCTION

Maintaining strict security and service-level objectives (SLOs) for modern disaggregated workloads (e.g., cloud computing, the internet of things, and augmented/virtual reality) demands that computationally intensive management and control decisions are made on the *current state of the entire datacenter network* (e.g., topology, queue sizes, and link and server loads) and applied *per-packet at line rate* [170]. A delay of even a few microseconds in today's petabit-bisection-bandwidth networks [60] could result in missing millions of anomalous packets [8, 17, 154], saturating switch queues and causing congestion [66, 162, 169], excessive retransmissions due to packet drops [37], or imbalanced traffic and server loads [4, 85]. However, current implementations face a dichotomy: they must choose between per-packet, line-rate execution or computational complexity.

Data-plane ASICs (e.g., switches and NICs) can react in nanoseconds to network conditions, but have a constrained programming model designed to forward packets at line rate (e.g., flow tables [15, 63]). Such a model restricts network operations to simple heuristics [4, 85, 98] or requires having purpose-built tasks in fixed-function hardware (e.g., middleboxes [29, 103]).

Control-plane servers can make complicated, data-driven decisions, but infrequently (typically, on each flow's first packet). The round trip (10 µs or more) between the controller and switch fundamentally limits the control plane's reaction speed, even with fast packet IO [38, 83, 132] and dedicated hardware (e.g., TPU [84] or GPU [119]). Moving computation to the switch CPU also does not help because it lacks global network knowledge, and the PCIe interface between the switch ASIC and CPU adds around 900 ns of round-trip delay [117]—at 12.8 Tb/s, the ASIC would have forwarded about 12k 128B packets (or 22k 64B packets) by the time a decision is made.

So, we ask the question: *"How can we delegate the complex decision-making of the control plane to a per-packet data-plane ASIC?"* The control plane can learn new trends (like attacks [106, 153], traffic patterns [162, 169], and workloads [169]) over time by sampling global network state and can train a machine-learning (ML) model to handle events of such types. The data plane can then use the model, which encapsulates network-wide behavior, to make forwarding decisions without visiting the control plane. Meanwhile, the control plane can continue to capture both switches' decisions and their impact on metrics (like flow-completion time), trains and optimizes the model to learn new event types (or signatures) as well as improve decision quality, and updates the switch model at fixed intervals. This training would happen at a coarser timescale (in tens of milliseconds), but off the critical path. And, using the most recently trained ML model, the data plane can make decisions starting at a flow's first packet while reacting to events of known (learned) types autonomously, requiring control-plane intervention only for new (unseen) event types.[1]

The challenge then is how to run the ML model in the data plane. We could map the output of the ML model as flow rules in the switch's match-action tables (MATs) [15] if the network behavior is stable. However, modern datacenter networks are dynamic, which would result in frequent table misses and control-plane visits, thus inflating the time it takes to react to network events [170]. Implementing models—specifically, deep neural networks (DNNs)—directly on existing switches is also not feasible [144, 168]. Most ML algorithms are built around linear algebra, which uses a significant amount of repetitive computation performed on a small number of weights with regular communication [88, 127, 150]. The match-action pipelines' VLIW architecture [15] lacks the necessary loops and multiplication operations; unnecessary flexibility, such as all-to-all VLIW communication [178], large memories, and ternary CAMs in MATs [15], consumes chip area without benefiting ML [144]. In short, existing data-planes lack the compute resources needed to execute modern ML algorithms (e.g., DNNs), while control-plane servers (and accelerators) are not optimized

[1] Flow rules, on the other hand, require control-plane intervention for every event, regardless of its type (or signature) [96].

for low-latency per-packet operations at network speeds exceeding multi-terabits per second.

In this paper, we propose *Taurus*, a domain-specific architecture for per-packet ML in data planes. Taurus extends the Protocol-Independent Switch Architecture (PISA) [15, 63] with a new compute block based on a parallel-patterns abstraction (i.e., MapReduce) [88], which supports data-parallelism common in modern ML applications [28]. The MapReduce block implements a spatial SIMD architecture, consisting of memory units (MU) and compute units (CU) interleaved in a grid and joined by a static interconnect (§4). Each CU has four pipelined compute stages, each performing an 8-bit fixed-point operation for each of the 16 independent data elements (lanes). When evaluating a 16-input perceptron, the CU uses the first stage to *map* 16 parallel multiplications; then, it uses the second stage to *reduce* the multiplied values into a single unit. A second CU applies an activation function (e.g., ReLU [112]). Multiple CUs can be used in parallel for hierarchical MapReduce computations (wide model layers) or in series for pipelined computations (multiple layers).

The MapReduce block works alongside parsers, MATs, and the scheduler to forward packets, with MATs connecting MapReduce to the pipeline: preprocessing MATs extract, format, and record packet-level, flow-level [154], cross-flow, and device features (using in-band network telemetry or INT [87]), the MapReduce block uses these features and an ML model to generate a numeric result, and post-processing MATs transform this output into a packet-forwarding decision. Packets that do not need an ML decision can bypass the MapReduce block, incurring no additional latency.

In summary, Taurus is an integrated system, bringing together ideas from network and ML architecture domains (i.e., extending PISA pipelines with a MapReduce unit) to enable a novel computing paradigm—*per-packet ML*—for the industry and research community to explore and innovate on.[2] We make the following contributions:

- A hardware design (§3) and implementation (§4) of a Taurus switch with a reconfigurable SIMD dataflow engine for MapReduce.
- Analysis of the ASIC[3] (§5.1.1) using a 15 nm predictive PDK [12] and synthesized against real ML networking applications (§5.1.2), microbenchmarks (§5.1.3), and MAT-only implementations (§5.1.4) to determine speed and area overheads relative to commercially available switches. Taurus's MapReduce block adds on average 122 ns latency to our line-rate (1 GPkt/s) ML models, while incurring an area and power overhead of 3.8% and 2.8%, respectively, for the largest block configuration.
- End-to-end system evaluation using a Taurus testbed based on a programmable switch connected to an FPGA that emulates our MapReduce block (§5.2.1). We show that Taurus achieves full model accuracy and detects two orders of magnitude more events than the control plane (§5.2.2). Furthermore, the data-plane model learns to handle new (anomalous) behaviors from the controller within milliseconds (§5.2.3).

[2]The source code for our Taurus prototype is publicly available at https://gitlab.com/dataplane-ai/taurus.

[3]We evaluate our design using Plasticine [127], a recent spatial SIMD accelerator, and modify it to target line-rate, inference-only applications. This entails lower precision, shorter pipelines, no floating-point support, no DRAM, and smaller on-chip memories.

## 2 THE NEED FOR PER-PACKET ML

To meet the strict SLOs of modern hyper-scale datacenters, the networking community is already running services (like load balancing [4, 85], anomaly detection [95, 102], and congestion control [162, 169]) in the network on a per-packet basis using programmable data planes (like Barefoot's Tofino chip [114, 115]). However, the constrained programming model of these data planes limits these services to simple heuristics, which cannot handle the complex interactions governing these massive datacenter networks [75]. Machine learning (ML), on the other hand, can approximate many of these complex interactions [43, 52]. And, through automated decision-making, ML algorithms exploit the vast quantity of data flowing within the network to learn progressively more informed policies [6, 17, 42, 101, 110, 123, 152, 153, 172–174]—tailored to a particular data center [99, 131, 169]. Moreover, recent work on ML for congestion control [162, 169], packet classification [99, 126, 131], and anomaly detection [106, 153] show that these algorithms have smoother and more accurate decision boundaries than manually-written heuristics. For example, ML provides a better throughput-delay tradeoff for congestion control compared to existing mechanisms (e.g., PCC [37] and QUIC [91]).

However, ML typically is used in a closed-loop decision-making system; hence, its reaction speed is critical to datacenter performance and security. For example, as shown in Table 1, active queue management (e.g., RED [45, 70] and DCTCP [5]) must make a mark/drop decision per-packet. Intrusion/fault detection and mitigation schemes (e.g., for heavy hitters [148], microbursts [139], DoS [17, 36]) and gray failures [77, 170]) must act quickly, as even a few microseconds' delay can miss millions of packets in a network with petabits-per-second of bisection bandwidth. These packets can have serious repercussions; for example, each missed packet during a SYN-flood attack will open a new connection on the server, unnecessarily consuming CPU and memory resources, as well as network bandwidth. Similarly, effective resource allocation (e.g., link bandwidth) requires timely classification of network traffic [11]. The consequences of slow ML decisions vary across these scenarios, but would generally result in sub-optimal behavior (i.e., missing security and service-level objectives)—the added latency results in delayed decisions and, ultimately, lower accuracy for the ML model (§5).

### 2.1 Limitations of Data-Plane ML

There have been a number of recent attempts to use current switch abstractions (i.e., MATs) [136, 144, 168] and specialized hardware [53] for in-network ML; however, both these approaches have deficiencies that prevent them from providing line-rate, per-packet ML.

**2.1.1 Inference on MATs.** The match-action abstraction, based on a VLIW architecture, is insufficient for line-rate ML in modern data-plane devices due to both missing operations (especially loops and multiplication) and inefficient MAT pipelines [15]. Binary neural networks have been implemented (using tens of MATs) but they are imprecise [136, 144]. Likewise, an SVM for IoT classification [168] is shown to consume eight additional tables on a NetFPGA reference switch—an experimental research platform [104, 113]. As a result, the resource usage is exorbitant relative to the installed model's quality (§5.1.4).

**Table 1: In-network applications demand fast reaction time (i.e., per-packet, -flowlet, -flow, or -µburst).**

| | Reaction Time | | | |
|---|---|---|---|---|
| **Application** | **Pkt** | **Flowlet** | **Flow** | **µburst** |
| **Security:** | | | | |
| - Heavy Hitters [148] | | | ✓ | |
| - DoS (e.g., SYN Flood) [159, 175] | ✓ | | ✓ | ✓ |
| - Probes (e.g., Port Scan) [154] | | | ✓ | |
| - U2R: Unauth. Access to Root [36] | ✓ | | | |
| - R2L: Unauth. Remote Access [36] | ✓ | | | |
| **Performance:** | | | | |
| - Congestion Control [82, 169] | ✓ | | | |
| - Active Queue Mgmt (AQM) [70] | ✓ | | | |
| - Traffic Classification [168] | ✓ | | ✓ | |
| - Load Balancing [85] | ✓ | ✓ | | |
| - Switching and Routing [163] | ✓ | | ✓ | |

**Table 2: Inference time for control-plane accelerators.**

| Accelerator | Latency (ms) |
|---|---|
| Broadwell Xeon | 0.67 |
| Tesla T4 GPU | 1.15 |
| Cloud TPU v2-8 | 3.51 |

***VLIW vs. SIMD Parallelism.*** Single-instruction/multiple-data (SIMD) has cheaper per-operation costs than switches' VLIW models. VLIW models, used in current switch MATs [15], have multiple logically-independent instructions per stage operating in parallel, reading and writing arbitrary locations in a packet-header vector (PHV). This all-to-multiple input communication and multiple-to-all output communication requires large crossbars and limits the number of instructions per stage. For example, a 16-issue VLIW processor has 20× as much control logic as an equally-powerful cluster of eight dual-issue processors [178]. Therefore, Barefoot's Tofino chip only executes 12 operations per stage: four of each of 8, 16, and 32 bits [65]. A typical DNN layer may require 72 multiplications and 144 additions [153]; even if multiplication were added to MATs, this would be 18 stages (most of them).

#### 2.1.2 Inference on Accelerators.

Traditional accelerators, like TPUs [84], GPUs [119], and FPGAs [44] could extend the data plane as bump-in-the-wire inference engines connected via PCIe or Ethernet. In most accelerators, inputs are batched to increase parallelism: larger batch sizes boost throughput by enabling more-efficient matrix-matrix multiplication. However, *unbatched* (matrix-vector) execution is needed for deterministic latency; otherwise, packets in the data plane would need to stall while waiting for a batch to fill. Moreover, adding another physically-separate accelerator would either consume switch ports (wasting transceivers and bandwidth) or replicate switch functions like packet parsing and match-action rules for feature extraction. Therefore, separate accelerators would add redundant area, decrease throughput, and consume power.

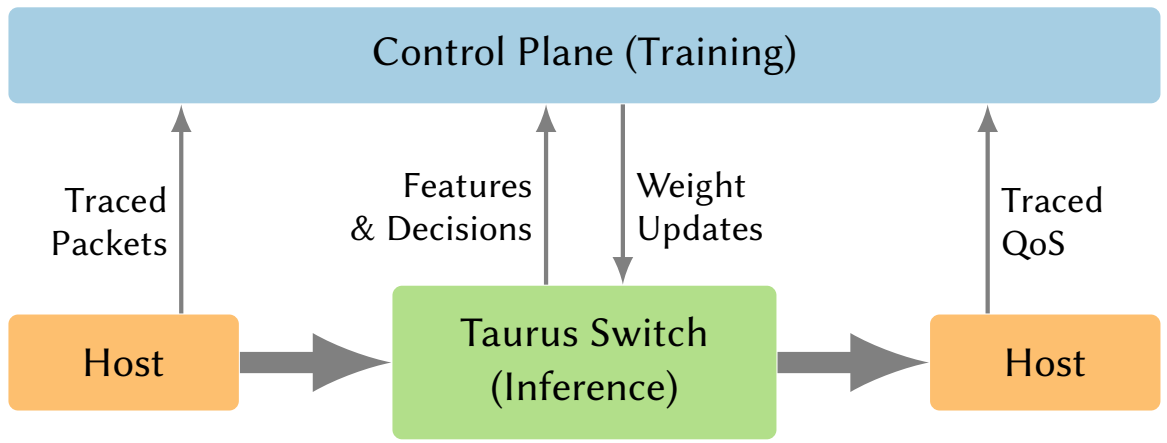

**Figure 1: Per-packet ML: training and inference. Hosts randomly mark packets to trace forwarding decisions and QoS metrics in the network, to update weights.**

### 2.2 Limitations of Control-Plane ML

Instead of executing an ML model per-packet, MATs [108] could cache inference results computed in the control plane. In a caching scheme, packets with previously-unseen features would be sent to the control plane for inferences, and the results would be stored in the data plane as flow rules. However, ML models with varying inputs, like packet size, would experience frequent and expensive trips to the control plane.

These cache misses have unavoidable penalties, due to the round-trip time for the control plane and software overhead—even with accelerators—that can hurt the accuracy of a model drastically. Table 2 benchmarks the latency of a DNN model for anomaly detection [153] on a vectorized CPU [1], a GPU [119], and a TPU [84] for unbatched inference. This latency comes from accelerator setup overhead (e.g., Tensorflow [1])—a CPU is the fastest design, but still takes 0.67 ms. Finally, rule installation time (3 ms for TCAMs [25]) would limit caching, especially, because it increases with flow-table size [47, 90].

## 3 TAURUS ARCHITECTURE

Taurus is a new data-plane architecture for switches (and NICs) that runs ML models at line rate for every packet and uses the model's output for forwarding decisions.

***High-Level System Overview.*** In a Taurus-enabled data center, the control plane gathers a global view of the network and trains ML models to optimize security and switch-level metrics. Meanwhile, the data plane uses these models to make per-packet, data-driven decisions. Unlike traditional SDN-based data centers, the control plane now installs both weights and flow rules into switches (Figure 1). Weights are more space-efficient than flow rules: for example, matching the behavior of our benchmark DNN (§5.1.2) may require 12 MB of flow rules (the full dataset), but only 5.6 KB of weights—a 2135× reduction in memory usage. Using monitoring frameworks like Deep Insight [78], the control plane can identify the impact of ML decisions and optimize weights accordingly.

***Anomaly Detection Case Study.*** Throughout this section, we use an ML-based anomaly detection (using a 4-layer DNN [153]) as a running example to motivate and describe the various logical components of the Taurus data-plane pipeline (Figure 2). As packets enter a switch, they are first parsed into Packet Header Vectors (PHVs) [15]—a fixed-layout, structured format—to extract header-level features (e.g., connection duration, bytes transferred, as well as protocol and service type). Next, switches use preprocessing MATs to look up domain-level features (e.g., matching IP addresses against autonomous systems' subnets to indicate ownership or geographic location)—discovered and installed by the control plane—and perform data integration and augmentation using VLIW actions on the header fields (§3.1). Taurus then uses postprocessing MATs to transform our ML model's output into a decision (§3.2) once the MapReduce block finishes executing the model on the extracted features (§3.3). This decision is then used to schedule, forward, or drop the packet. We share existing MATs in PISA-based switches for pre/postprocessing.

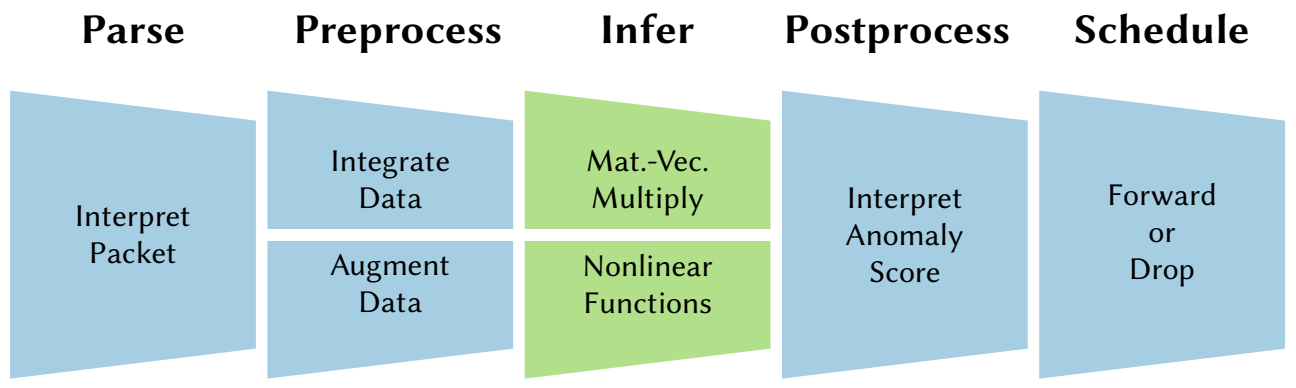

**Figure 2: The logical steps for an anomaly-detection application in Taurus.**

## 3.1 Parsing & Preprocessing

Before inference, Taurus processes raw packet headers into canonical form through MATs, adding or repairing packet-level data as needed. We use stateful elements (i.e., registers) of the switch-processing pipeline to aggregate features across packets and across flows. MATs then add these aggregates to each packet's metadata to enhance per-packet predictions. Data preprocessing can also use MATs to convert header fields into features for the ML model. For example, in our anomaly-detection example, MATs would format features as fixed-point numbers and the MapReduce block would decide if a packet is anomalous or benign.

Taurus replaces categorical relationships with simpler numeric relationships using lookup tables; for example, a table transforms port numbers into a linear likelihood value, which is easier to infer from [31]. Preprocessing can also invert the probability distribution underlying a sampled value. Taking a logarithm of an exponentially distributed variable results in a uniform distribution, which an ML model can process with fewer layers [138]. Such *feature engineering* transfers the load from an ML model to its designer; refining features can increase accuracy with constant size [16, 138].

Finally, in-band network telemetry (INT)—measurements embedded into packets—provides switches with a view of global network state [87]. As a result, Taurus devices are not limited to inference using switches' local state. Instead, models can examine the packet's entire history, through INT, and the flow's entire history, through stateful registers, to increase their predictive power (e.g., counting urgent flags across a flow or monitoring connection duration).

## 3.2 Postprocessing & Scheduling

MATs also interpret ML decisions. For example, if our anomaly-detection model outputs 0.9 (suggesting a likely-anomalous packet), MATs decide how to handle the packet, either dropping, flagging, or quarantining it. In Taurus, these postprocessing MATs connect inference to scheduling, which uses abstractions like PIFO [147] to support a variety of scheduling algorithms.

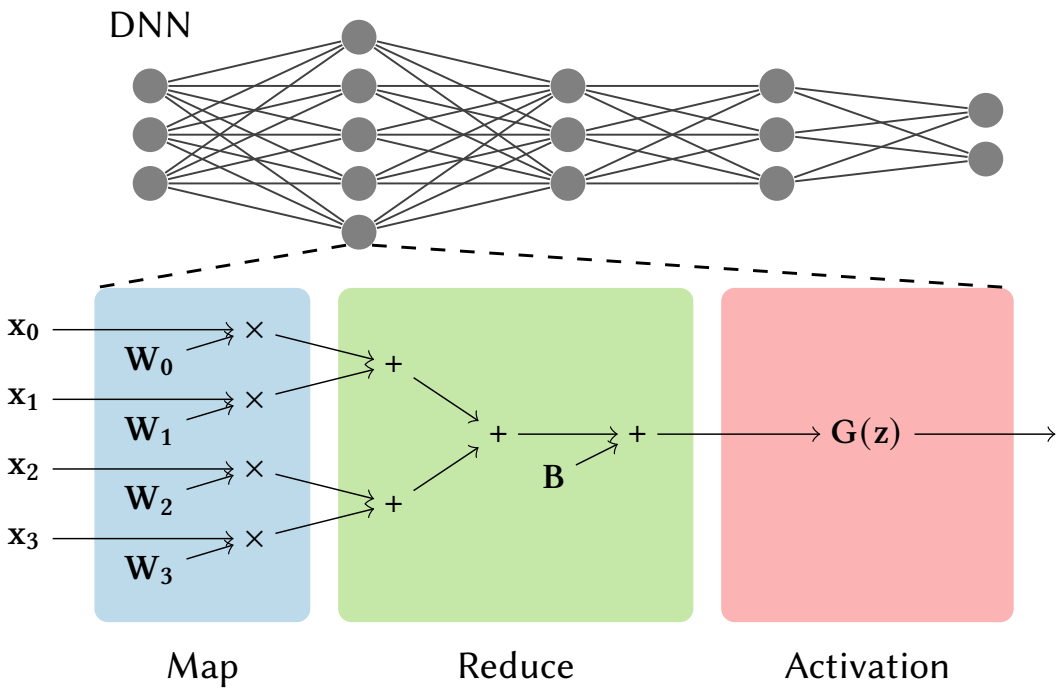

**Figure 3: The compute graph of a perceptron with the breakdown between map, reduce, and activation functions (outer-loop map). The perceptron is a building block of a larger DNN.**

ML models will provide probabilistic guarantees, but we can constrain their behavior with hard bounds to ensure robust network operation. The control plane can compile high-level *safety* (no incorrect behavior) and *liveness* (eventual correct behavior) properties into per-switch constraints as postprocessing flow rules. By constraining the ML model's decision boundary, the data plane can guarantee correct network behavior without complicated model verification.

## 3.3 MapReduce for Per-Packet ML Inference

For each packet, inference combines cleaned features and model weights to make a decision. ML algorithms, like support-vector machines (SVMs) and neural networks, use matrix-vector linear algebra operations and element-wise non-linear operations [59, 72]. Non-linear operations let models learn non-linear semantics; otherwise, the output would be a linear combination of the inputs. Unlike header processing, ML computation is very regular, using many multiply-add operations. In the more computationally-taxing linear portion of a single DNN neuron, input features are each multiplied by a weight, then added to yield a scalar value. Generalizing this operation, vector-to-vector (*map*) and vector-to-scalar (*reduce*) operations suffice for the computationally-intensive linear portions of a neuron. This, along with the limitations of current switch architectures, motivates the need for a new data-plane abstraction, *MapReduce*, that is flexible enough to express a variety of ML models but specific enough to allow efficient hardware development.

**3.3.1 The MapReduce Abstraction.** Our design uses MapReduce SIMD parallelism to provide high computational throughout, cheaply. *Map* operations are element-wise vector operations, such as addition, multiplication, or non-linear operations. *Reduce* operations combine a vector of elements to a scalar value using associative operations like addition or multiplication. Figure 3 shows how map and reduce are used to compute a single neuron (dot product), which can be combined hierarchically into large neural networks.

```
Control Parser (...) {...}
Control PreProcessMAT (...) {...}
Control MapReduce( inout metadata FeatureSet,
                   inout metadata Output ) {
  Weights = loadModelFromFile(Anomaly.model)
  LinearResults = Map(sizeOf(Weights[0])) { i =>
    Mult_Results = Map(sizeOf(Weights[1])) { j =>
      Weights[i,j] * FeatureSet[j] }
    Reduce(Mult_Results) { (x,y) => x + y } }
  Output = Map(sizeOf(LinearResults)) { k =>
    ReLU(LinearResults[k])
  } }
Control PostProcessMAT (...) {...}
Control Deparser (...) {...}
```

**Figure 4: MapReduce syntax in P4 for a DNN layer of our anomaly-detection example, based on Spatial [88].**

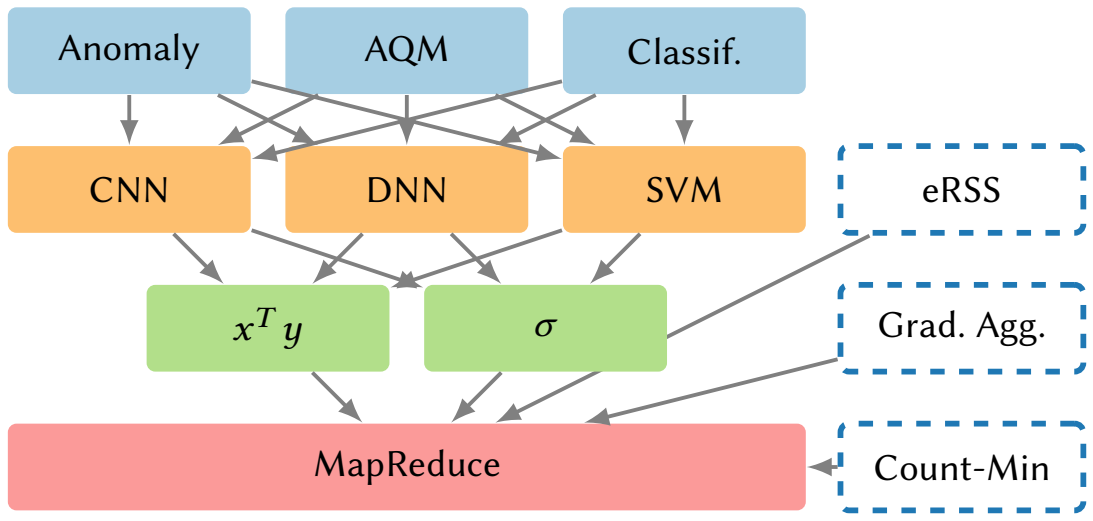

**Figure 5: ML applications (top) map to models and simpler primitives, which compile to MapReduce. Additional applications (right) map directly to MapReduce.**

MapReduce is a popular form for ML models: MapReduce can accelerate ML both in distributed systems [22, 54, 55, 57, 133] and at a finer granularity [20, 21, 28, 150].

***A MapReduce Control Block in P4.*** To program Taurus, we introduce a new dedicated control-block type in P4 [13, 122], called `MapReduce`, in addition to the ones used for ingress and egress match-action computations. Within this new control block, MapReduce units can be invoked using the `Map` and `Reduce` constructs. Figure 4 shows our proposed MapReduce syntax—inspired by the recently proposed Spatial language [88]—for the anomaly-detection example. The outermost map iterates over all the layer's neurons, while the inner MapReduce pair performs the linear operation for each neuron. A final map instruction applies an activation function (i.e., ReLUs or sigmoids). Other than `Map` and `Reduce`, the only additional constructs needed are arrays and out-of-band weight updates.

**3.3.2 Broader Application Support.** By providing common primitives, we can support a set of applications broader than ML (Figure 5), including stream processing for data analysis [18, 89], gradient aggregation for large-scale distributed training [57, 97, 111, 137], and more at the switch and NIC [125]. For example, Elastic RSS (eRSS) uses MapReduce for consistent hashing to schedule packets and cores: map evaluates cores' suitability, and reduce selects the closest core [134]. MapReduce can also support sketching algorithms, including Count-Min-Sketches (CMS) [30] for flow-size estimation. Furthermore, recent research shows that Bloom filters can also benefit from, or be replaced by, neural networks [130]. In essence, Taurus provides a programmable data-plane abstraction (MapReduce) that can support a large class of applications more efficiently (in terms of hardware resource usage and performance, §5), compared to existing data-plane abstractions (i.e., MATs) alone.

## 4 TAURUS IMPLEMENTATION

The complete physical data-plane pipeline of a Taurus device is shown in Figure 6, consisting of control blocks for packet parsing, ML with MapReduce, packet forwarding with MATs, and scheduling as well as a bypass path for non-ML packets. Taurus's packet parser, pre/postprocessing MATs, and scheduler use existing hardware implementations [15, 56, 147]. We base Taurus's MapReduce block on Plasticine [127], a coarse-grained reconfigurable array (CGRA) composed of a sea of compute and memory units, which are reconfigurable to match applications' dataflow graphs. However, Plasticine was originally used for designing standalone accelerators, while we need a latency-optimized streaming fabric to operate in the network.

***MapReduce: Compute Unit (CU).*** Each compute unit (CU, Figure 8) is composed of functional units (FUs) organized in *lanes* and *stages* and performs a map, a reduction, or both. Within a CU stage, all lanes execute the same instruction and read the same relative location. CUs have pipeline registers between stages, so every FU is active on every cycle; pipelining also occurs at a higher level between CUs.

Because the fabric needs to operate as part of a full switch ASIC, resource efficiency is key. We use fixed-point reduced precision hardware to execute the arithmetic needed for the linear algebra in ML algorithms. Fixed-point hardware is faster, more area-efficient, and consumes less power than floating-point operations. Furthermore, we customize the lane-to-stage ratio in the CUs to fit the minimum requirements of our application space. A complete explanation of the design-space exploration process appears in §5.

***MapReduce: Memory Unit (MU).*** Next, we focus on the speed of memory accesses. Weights in an ML model need to be retrieved quickly if we want to make a decision on a per-packet level. SRAM-based operations can be done with single-cycle accesses, so we exclusively use on-chip memories. DRAM controllers are eliminated, because accesses take on the order of 100s of cycles. While this reduces the size of the models that Taurus can support, at a 1 GHz clock it ensures nanosecond-level latencies. We use banked SRAMs as memory units (MUs), which are interspersed with CUs in a checkerboard pattern for locality, to store the weights of ML models (Figure 7). This also allows coarse-grained pipelining, where CUs perform operations and MUs act as pipeline registers.

Multiple levels of pipelining within each CU and pipelining in the interconnect guarantee a 1 GHz clock frequency—a crucial factor for matching the line rate of high-end switch hardware [15, 147].[4]

***Pre/Postprocessing MATs.*** By using MATs (VLIW) for data cleaning and MapReduce (SIMD) for inference, Taurus combines different

[4]The CU and MU architecture in Taurus currently supports dense ML models. However, it can be extended to support sparse linear algebra [135], which we leave as future work.

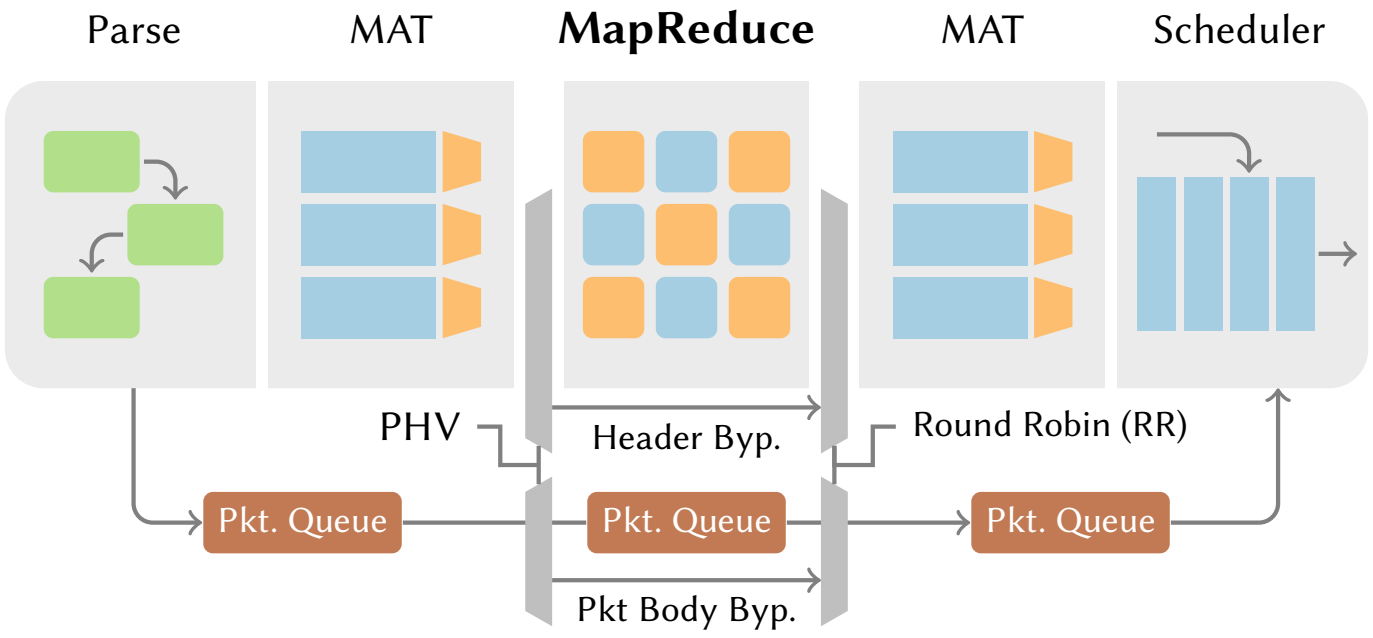

**Figure 6: Taurus's modified data-plane pipeline, including bypass paths for non-ML packets. A preprocessing MAT decides whether to bypass ML, as metadata in the PHV, and a round-robin (RR) selector arbitrates which path to connect to the postprocessing MAT.**

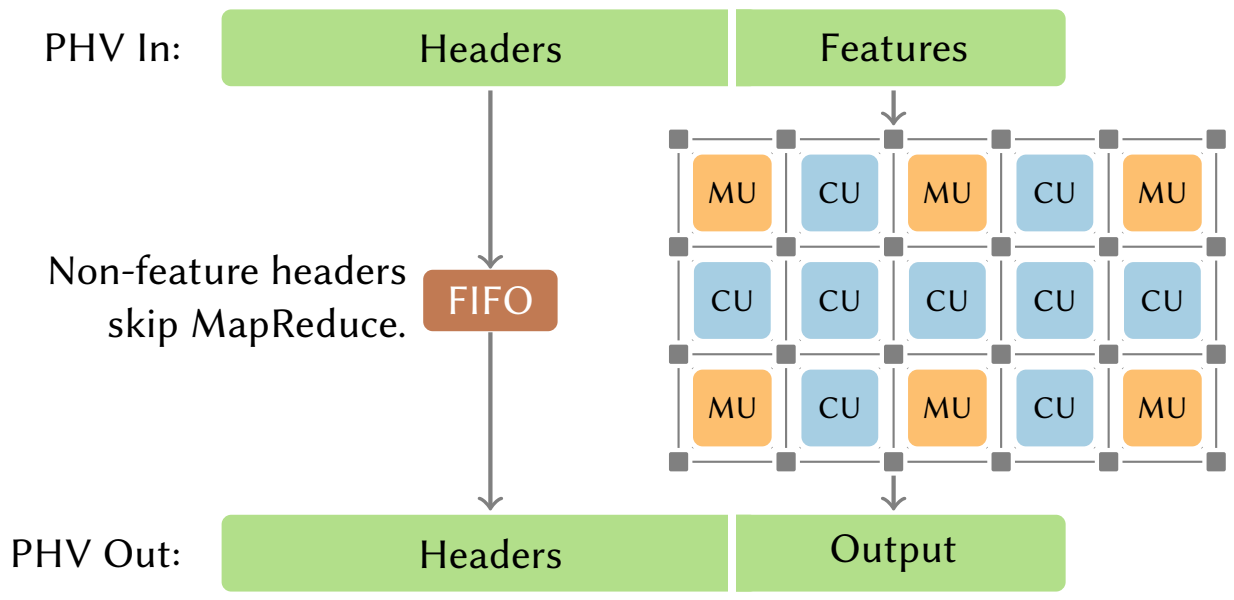

**Figure 7: Taurus's MapReduce block and its interface to the rest of the pipeline.**

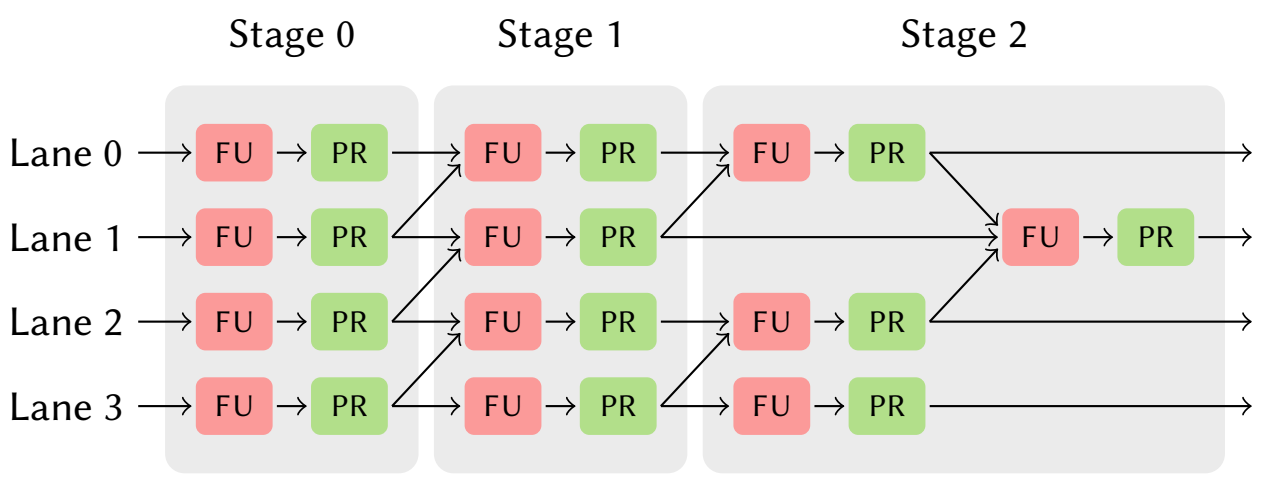

**Figure 8: A three-stage CU, composed of functional units (FUs) and pipeline registers (PRs). The third stage supports map and sparse reductions.**

models of parallelism to build a fast and flexible data-plane pipeline. MATs are connected to the MapReduce block through the same PHV interface, which is used to connect other stages in the pipeline. Only a fraction of the PHV, containing features, enters the MapReduce block, while other headers progress directly to the postprocessing MATs, as shown in Figure 7.

***Non-ML Traffic Bypass.*** For packets that do not require ML inference, Taurus forwards them directly to the postprocessing MATs, bypassing MapReduce (Figure 6). It splits traditional switches' single, large packet queue [15] into three sub-queues: one for each of the preprocessing MATs, MapReduce block, and postprocessing MATs; allocating it proportionally across these blocks based on their pipeline depth. Preprocessing MATs forward ML packets to the MapReduce block—the PHV progresses through the block and body is enqueued into the corresponding queue—and non-ML packets are sent directly to the postprocessing MATs, without incurring any additional latencies. Moreover, non-feature headers of a packet also follow the bypass path to the postprocessing MATs, and only the required feature headers enter the MapReduce block as a dense PHV (to minimize sparse data occurrences).

***Target-Independent Optimizations.*** MapReduce is general enough to support target-independent optimizations: optimizations that consider available execution resources (parallelization factors, bandwidth, and more) without considering hardware-specific design details [128, 176]. Parallelizing MapReduce programs unrolls loops in space: if sufficient hardware resources are available, a model can execute one iteration per cycle. As loop unrolling happens at compile-time, Taurus can guarantee deterministic throughput: either line-rate performance, or some known fraction thereof. Static line-rate reduction already occurs in switches via recirculation [15] and in datacenter networks through link oversubscription [62, 118].

Latency, in addition to area, also limits switch-level ML because switches must forward packets in hundreds of nanoseconds. Latency increases with depth, so datacenter SLOs would effectively limit models' layer counts. By preprocessing features with MATs, we can provide adequate accuracy with fewer layers and less latency: the model must only learn inter-feature relationships, not the header-feature mapping.

***Target-Dependent Compilation.*** A variety of programming languages natively support MapReduce [71, 109, 121, 155]. To support our SIMD-based fabric, we program Taurus's MapReduce block using a modified version of Spatial [88], a domain-specific language (DSL) based on parallel-patterns that represents MapReduce programs as a sequence of nested loops. It supports target-dependent optimizations as well as target-independent optimizations for Taurus. Programs are compiled to a streaming dataflow graph: from this hierarchy, innermost loops become SIMD operations within a CU, and outer loops are mapped over multiple CUs. Then, overly-large patterns (those requiring too many compute stages, inputs, or memory banks) are split into smaller patterns that fit in CUs and MUs; this is necessary to map non-linear functions with long basic blocks (i.e., long sequences of code with no branches). Finally, the resulting graph is placed and routed on the MapReduce block's interconnect.

# 5 EVALUATION

## 5.1 Taurus ASIC Analysis

We first examine our MapReduce block by analyzing its power and area (§5.1.1). We then evaluate its performance by compiling several recently-proposed networking ML applications (§5.1.2). Next, we demonstrate its flexibility using common ML components, which can be composed to express a variety of algorithms (§5.1.3). Finally, we compare it with existing MAT-only ML implementations (§5.1.4).

**Table 3: Accuracy of DNNs for TMC IoT traffic classifiers [145]; this shows minimal loss for 8-bit quantization.**

| DNN Kernel | Model Accuracy (%) | | |
|---|---|---|---|
| | `float32` | `fix8` | Diff. |
| $4 \times 10 \times 2$ | 67.06 | 67.01 | -0.05 |
| $4 \times 5 \times 5 \times 2$ | 67.02 | 66.95 | -0.07 |
| $4 \times 10 \times 10 \times 2$ | 67.04 | 67.02 | -0.02 |

**Table 4: Area and power scaling (per-FU) at the target design (16 lanes, 4 stages) for different precisions. (Bars represent each entry as a fraction of the largest entry in that column.)**

| Precision | Area ($\mu m^2$) | Power ($\mu W$) |
|---|---|---|
| `fix8` | 670 | 456 |
| `fix16` | 1338 | 887 |
| `fix32` | 2949 | 2341 |

#### 5.1.1 Design Space Exploration.

Taurus's MapReduce block is parameterized, including precision, lane count, and stage count. To estimate Taurus's area and power, we use ASIC synthesis and FreePDK15, a predictive 15 nm standard cell library [12]; we also use CACTI 7.0 [9] to estimate memory bank area. To guide our evaluation, we minimize area, power, and latency while targeting full model accuracy (i.e., no quantization loss) and per-packet throughput.

***Fixed-Point Precision.*** For ML inference, fixed-point arithmetic is faster than floating point with equivalent accuracy [67, 84]. We use 8-bit precision for Taurus, because it is shown to be sufficient for inference (compressed models use even fewer bits) [34, 100, 160]. In Table 3, we can see that for a variety of DNNs, the quantization error (using TensorFlow Lite [61, 81]) is negligible. The minimal accuracy loss and 4× reduction in resources (Table 4) justifies our reduced precision architecture.

***Lane Count.*** Ideally, the number of lanes in a CU would be matched to the amount of available vector (SIMD) parallelism. If too few lanes are provisioned, a vector operation would have to be mapped across multiple CUs, increasing the amount of control logic required and decreasing efficiency. Similarly, provisioning more lanes than the amount of vector parallelism would cause some to be unused: the CU can execute operations on only a single vector. Figure 9 shows that raw area efficiency (area per FU) increases with the number of lanes.

The anomaly-detection DNN [153] is our largest model requiring line-rate operation, so we use it to set the ideal lane count. The DNN's largest layer has 12 hidden units, so the largest dot-product calculations involve 12 elements; the 16-lane configuration fully unrolls the dot product within a single CU while minimizing underutilization. Currently, the 16-lane configuration balances area overhead, power, and mapping efficiency, but optimal lane counts may change as data-plane ML models evolve. Because MapReduce

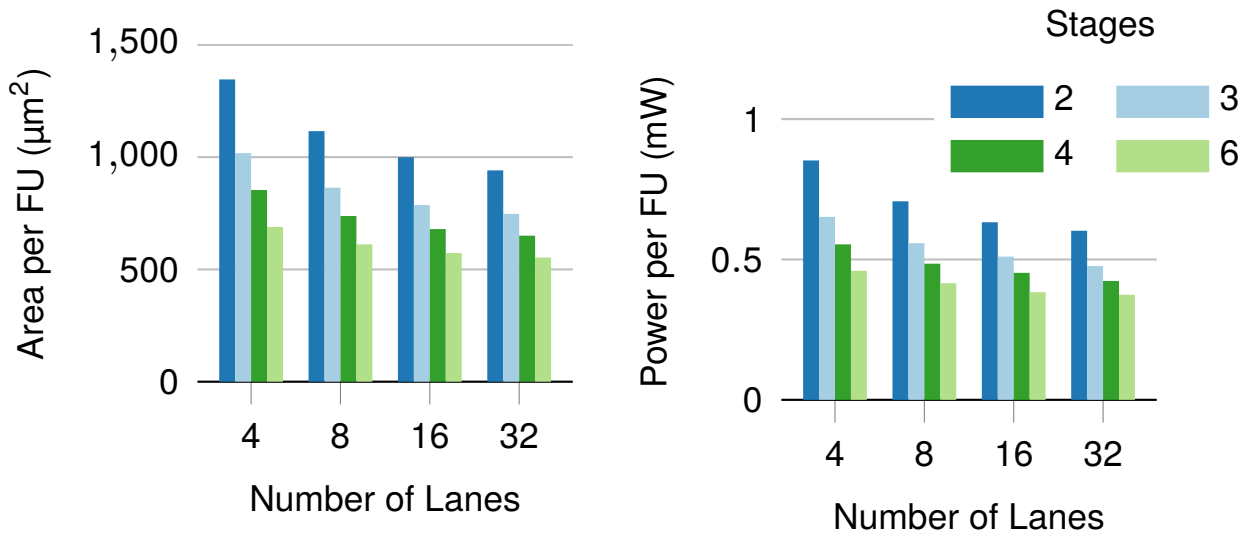

**(a) Area.** **(b) Power (10 % switching).**

**Figure 9: Area and power consumption per-FU for various CU configurations (lanes and stages).**

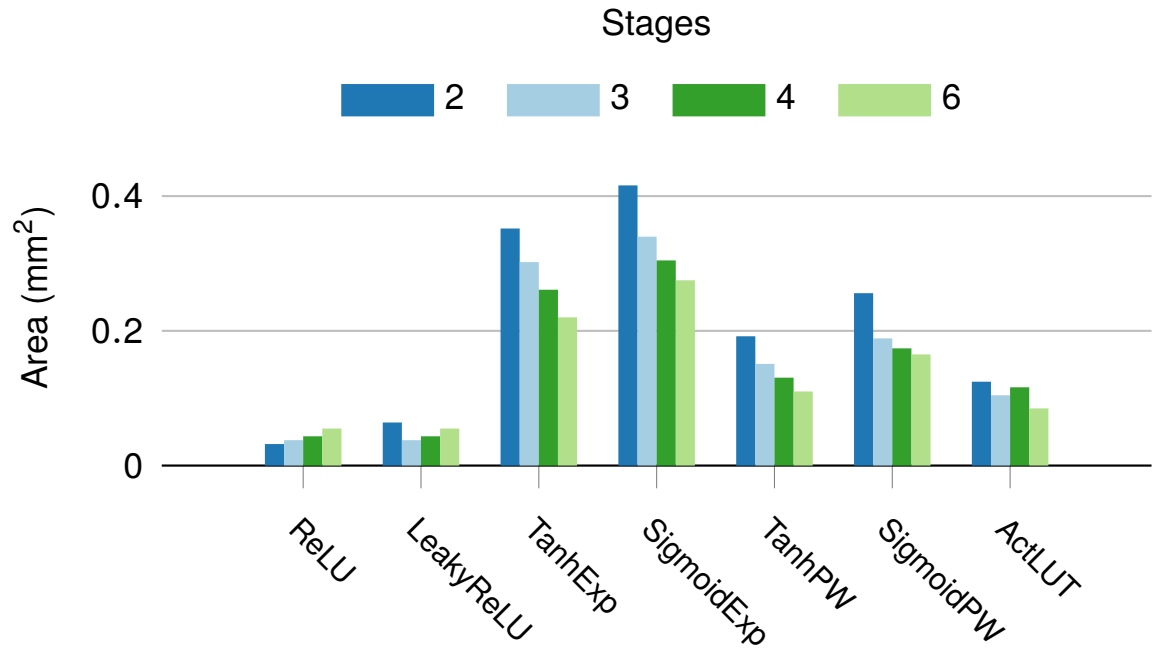

**Figure 10: Area needed for activation functions as the number of stages varies, all operating at line rate (1 GPkt/s).**

programs are hardware agnostic, the compiler will handle the differences in unrolling factors as needed (i.e., parallelism within a CU vs. across CUs).

***Stage Count.*** We perform a similar study to quantify the CU's stage count's impact on accuracy. Inner product, the linear operation behind most of our models, uses two stages: one map (multiplication) and one reduce (addition), while non-linear operations use a sequence of maps; therefore, we study stage count's impact on non-linear operations. Figure 10 shows the total area (the product of CU count and area) needed to implement a variety of activation functions using CUs of different depths. Furthermore, for shallow activation functions (e.g., ReLU), the later stages are not mapped, leading to area increases with more stages. Theoretically, more stages are more efficient (Figure 9) and with context merging can allow a complier to map even more complex activation functions [177]. However, our inner product and ReLU microbenchmarks—which form the core of many common networks—benefit from only two stages. Therefore, we pick four compute stages in our final ASIC design to support both.

***Final ASIC Configuration.*** Our final CU has 16 lanes, four stages, and an 8-bit fixed-point data path. Including routing resources [176], it takes 0.044 $mm^2$ (680 $\mu m^2$ per FU, on average). Each MU has 16 banks with 1024 entries each, and consumes 0.029 $mm^2$ including routing resources. Overall, we provision a $12 \times 10$ grid with a 3:1 ratio of CUs to MUs, taking 4.8 mm². Considering a switch with four reconfigurable pipelines having 32 MATs each, 50% of the chip area is taken up by the MATs [86]. And, the addition of a MapReduce block per pipeline increases the total chip area by 3.8%; an iso-area design would lose 3 MATs per pipeline. This is a negligible overhead in comparison to the types of programs we can support (§5.1.2). Taurus's ASIC parameters are based on the applications and functions in use today: with new models, new parameters may provide greater efficiency.

**Table 5: Performance and resource overheads of several application models. Overheads are calculated relative to a 500 mm² chip with 4 reconfigurable pipelines [65], with the system drawing an estimated 270 W while running at 1 GPkt/s line rate [3, 19, 114].**

| | | Perf. | | Area | | Power | |
|---|---|---|---|---|---|---|---|
| **App** | **Model** | **GPkt/s** | **ns** | **mm²** | **+%** | **mW** | **+%** |
| IoT | KMeans | 1.00 | 61 | 0.3 | 0.2 | 177 | 0.3 |
| Anom. | SVM | 1.00 | 83 | 0.6 | 0.5 | 395 | 0.6 |
| Anom. | DNN | 1.00 | 221 | 1.0 | 0.8 | 647 | 1.0 |
| Indigo | LSTM | — | 805 | 3.0 | 2.4 | 1897 | 2.8 |
| 12×10 Grid | | | | 4.8 | 3.8 | | |

**5.1.2 Application Benchmarks.** We evaluate Taurus using four ML models [106, 153, 168, 169]. The first, an IoT traffic classification, implements KMeans clustering using 11 features and five categories. The first anomaly-detection algorithm is an SVM [106] with eight input features selected from the KDD dataset [2, 36] and a radial-basis function to model nonlinear relationships. Our second anomaly-detection algorithm is a DNN that takes six input features (also a KDD subset); it has layers with 12, 6, and 3 hidden units [153]. Finally, the online congestion-control algorithm (Indigo [169]) is an LSTM. Indigo uses 32 LSTM units followed by a softmax layer and is designed to run at an end-host NIC. Although Indigo does not run per-packet, its update interval is significantly lower with Taurus: permitting more accurate control decisions and faster reaction times.

***Area & Power.*** We show area and power relative to an existing programmable switch ASIC[5] with a PISA pipeline [15] in Table 5, considering only the number of CUs and MUs performing useful work. Therefore, the actual area of a prototype for these benchmarks is the area of the largest benchmark, with unused CUs disabled for smaller benchmarks. Simple models, like SVM-based anomaly detection, have as little as 0.2% area overhead and 0.3% power overhead. Therefore, we choose Taurus's MapReduce block area as 4.8 mm², which is similar to recently-proposed switch additions (e.g., CONGA [4] and Banzai [146] consuming an additional 2% and 12% area, respectively). If only smaller models were supported, KMeans, SVMs, and DNNs would add only about 0.8% more area and 0.9% more power. For 16- and 32-bit data paths, both area and power will increase by about a factor of 2 and 4, respectively.

[5] With TSMC's reticle limit of 858 mm² [142], a 64×100 Gb/s switch chip today has a die size of 500–600 mm². (Source: private correspondence with switch chip vendors.)

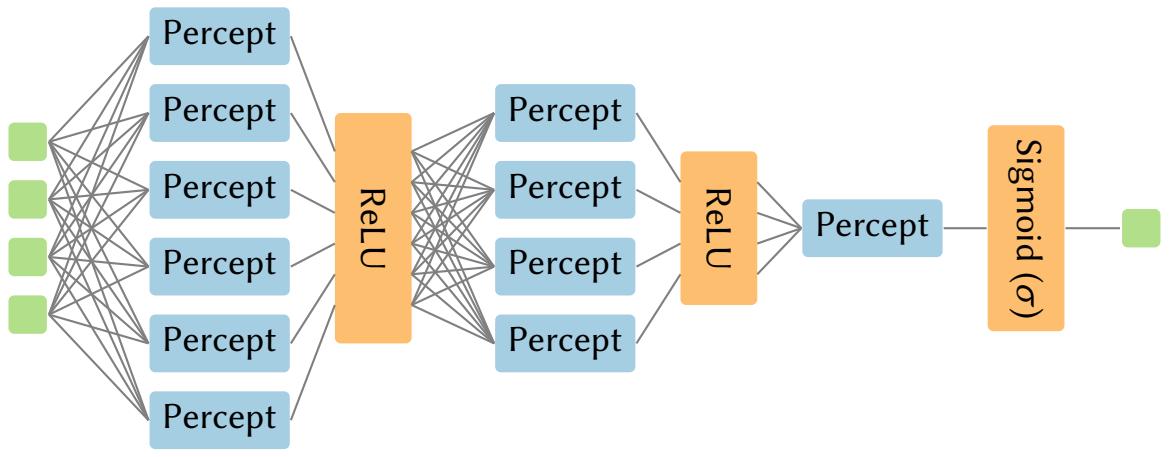

**Figure 11: A small DNN, broken down into independent microbenchmarks.**

***Latency & Throughput.*** KMeans, the SVM, and the DNN process one packet header per cycle (line-rate), and their latency remains in the nanosecond range (Table 5). Assuming a datacenter switch latency of 1 µs [35], KMeans, the SVM, and the DNN add 6.1%, 8.3%, and 22.1% more latency, respectively. In software, the Indigo LSTM significantly improves application-level throughput and latency [169], running every 10 ms—likely limited by the LSTM's computational requirements. In Taurus, Indigo can produce a decision every 805 ns: this allows the LSTM network to react more quickly to changes in load and better control tail latency. Taurus's bypass path also avoids adding latency for non-ML packets.

**5.1.3 Microbenchmarks.** Smaller dataflow programs can be composed into a single, large program: for example, Figure 11 shows a DNN built from several (linear) perceptron layers fused with non-linear activation functions. These microbenchmarks are general building blocks intended to show the versatility of a programmable, MapReduce-based fabric. Linear functions include a reduction network that limits the degree of communication-free parallelism. Conversely, nonlinear functions can be perfectly SIMD-parallelized because there is no interaction between adjacent data elements. For example, if the output of 16 different perceptrons is input to a ReLU, we simply map the ReLU over the 16 outputs, which are then computed in parallel. Table 6 shows the area and latency required for each microbenchmark when unrolled to run at line rate.

***Linear Operations.*** Our linear microbenchmarks are comprised of a one-dimensional convolution with eight outputs and a kernel dimension of two (frequently used to find spatial or temporal correlations [94]) and a 16-element inner product, which forms the core of perceptron neural networks, LSTMs, and SVMs. Because the convolution does not map well to vectorized MapReduce (there are multiple small inner reductions), it requires 8× unrolling and much chip area. However, the inner product runs at line rate in only a single CU; it can be efficiently composed into high-performance deep neural networks. The minimum latency for a 16-lane CU to perform a MapReduce is five cycles: one cycle for map and four cycles for reduce, using different fractions of a single stage for each reduction cycle (Figure 8). The remaining latency comes from data movement from the input to the CU and then to the output; Taurus takes roughly five cycles for each data movement—a result of spatially-distributed dataflow.

***Unrolling.*** Using MapReduce's target-independent optimization, large ML models can run over multiple cycles with a corresponding line-rate reduction (Table 7). Unrolling inner and outer loops will result in higher throughput and lower latency while consuming more area. However, not all benchmarks can have their outer loop unrolled: for example, the inner product has no outer loop. The iterative (i.e., loop-based) convolution runs at one-eighth of line rate, and unrolling it to meet line rate results in an 8× area increase.

**Table 6: Area and latency of each microbenchmark, running at line rate in a 16-lane, four-stage CU. (Latency is the sum of the perceptron and ReLU execution time.)**

| | µbmark | Area ($mm^2$) | Lat. (ns) |
|---|---|---|---|
| **Linear** | Conv1D | 1.57 | 122 |
| | Inner Product | 0.04 | 23 |
| **Nonlinear** | ReLU | 0.04 | 22 |
| | LeakyReLU | 0.04 | 22 |
| | TanhExp | 0.26 | 69 |
| | SigmoidExp | 0.31 | 73 |
| | TanhPW | 0.13 | 38 |
| | SigmoidPW | 0.17 | 46 |
| | ActLUT | 0.12 | 36 |

**Table 7: Throughput and area scaling of microbenchmarks with unrolling factors from 1 to 8.**

| µbmark | Unroll | Line Rate | Area ($mm^2$) |
|---|---|---|---|
| **Conv1D** | 1 | ⅛ | 0.19 |
| | 2 | ¼ | 0.44 |
| | 4 | ½ | 0.93 |
| | 8 | 1 | 1.57 |
| **Inner Product** | – | 1 | 0.04 |

***Nonlinear Operations.*** Activation functions are necessary to learn nonlinear behavior; otherwise, the entire neural network would collapse into a single linear function. Each activation function serves a different purpose: LSTMs use tanh for gating [73], while DNNs use the simpler ReLU and Leaky ReLU [112]. The most efficient functions (ReLU and Leaky ReLU) do not use lookup tables (LUTs) and take only a single CU. More complicated functions, including sigmoid and tanh, have several versions: Taylor series, piecewise approximations, and LUTs [67, 161], with Taylor series and piecewise approximations requiring 2–5 times as much area. LUT-based functions need memory: each table approximates activation functions by storing pre-computed output values as 1024 8-bit entries [67, 161]—a small fixed fraction of switch memory, even when replicated.

#### 5.1.4 Comparisons with MAT-Only ML Designs.

As discussed in §5.1.1, our final Taurus ASIC configuration—with 4 pipelines, each having a MapReduce block—consumes 3.8% additional chip area (or an iso-area equivalent of 3 MATs per MapReduce block). In comparison, recent MAT-only neural network (NN) implementations [144, 168] consumes 10s of MATs. For example, N2Net, a binary neural network (BNN) implementation, requires at least 12 MATs per layer [144]. This would take 48 MATs to support the anomaly-detection DNN [153]—Taurus consumes only 3. Similarly, the IIsy framework [168] implements non-NN algorithms on MATs, consuming 8 and 2 MATs for SVM and KMeans, respectively. A corresponding Taurus ASIC would incur 0.5% chip area (or a single MAT), only. Both N2Net and IIsy provide unique ways of mapping ML algorithms to MATs; however, for data-plane ML to become ubiquitous we need more efficient ML hardware for networks.

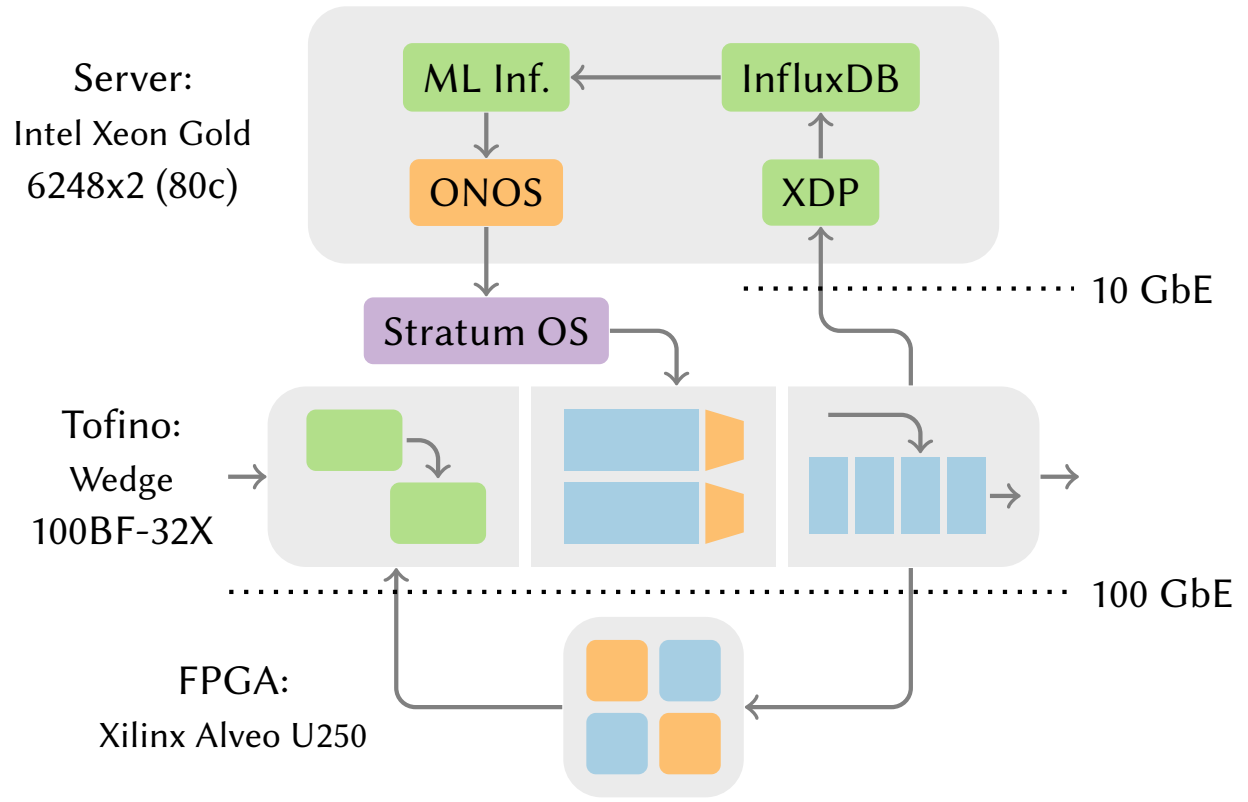

**Figure 12: Taurus testbed for inference: The server runs leading open-source software, i.e., ONOS [46], XDP [129], InfluxDB [33], and control-plane inference using TensorFlow [1]. The Tofino switch and FPGA implement Taurus's MAT pipeline and the MapReduce block, respectively.**

## 5.2 End-To-End Performance

#### 5.2.1 Taurus Testbed.

To evaluate Taurus's end-to-end performance, we build a testbed using industry-standard SDN tools, a Tofino switch, and an FPGA that implements the MapReduce block, as shown in Figure 12. With this testbed, we show that Taurus decides on a per-packet basis, is faster than conventional control-plane ML solutions, and has greater accuracy (§5.2.2).

***Control-Plane ML: Baseline.*** In our baseline,[6] the control-plane server samples telemetry packets through a 10 Gbps link using an XDP-enabled Intel X710 NIC [79] (running a custom XDP/eBPF program, 84 LoC), and stores them in an InfluxDB streaming database [33]. Our vectorized ML model (written in Keras [64], 272 LoC) runs inference on these packets in batches, and the Open Network Operating System (ONOS) [46] installs the model's outcome as flow rules on the switch.

***Data-Plane ML: Taurus.*** In Taurus,[6] a programmable Barefoot Wedge 100BF-32X [116] switch, running the Stratum OS [48] implements Taurus's PISA components. The match-action pipeline of the Tofino switch implements Taurus's parsing and preprocessing MATs (172 LoC). A 100 Gbps Ethernet link connects the switch to a Xilinx Alveo U250 FPGA [164], which emulates the MapReduce hardware; we compile our ML models to the FPGA using Spatial [88] (105 LoC) and the Xilinx OpenNIC Shell [166]. To stream packets, we integrate the Xilinx 100G CMAC [167] with an AXI stream interface [165] (298 LoC) to the MapReduce block in the FPGA. Once processed by the FPGA, the packets go through the postprocessing MATs in the switch before being forwarded to the network. Finally, we use an additional two 80-core Intel Xeon servers running MoonGen [39] to generate and receive traffic.

[6] XDP 2.6.32 | Tensorflow/Keras 2.4.0 | InfluxDB 1.7.4 | ONOS 2.2.2 | Stratum/Barefoot SDE 9.2.0 | Xilinx Vivado 2020.2 | Spatial `40d182` | MoonGen `525d991`

Table 8: Baseline ML's batch sizes and latency, and accuracy. Taurus detects two orders of magnitude more events.

| | Batch Size | | Baseline Latency (ms) | | | | | Detected (%) | | F1 Score | |
|---|---|---|---|---|---|---|---|---|---|---|---|
| Sampling | XDP | Rem. | XDP | DB | ML | Install | All | Baseline | Taurus | Baseline | Taurus |
| $10^{-5}$ | 1 | 5 | 3 | 14 | 16 | 2 | 34 | 0.781 | 58.2 | 1.549 | 71.1 |
| $10^{-4}$ | 2 | 33 | 2 | 17 | 18 | 4 | 41 | 2.553 | 58.2 | 4.944 | 71.1 |
| $10^{-3}$ | 17 | 637 | 3 | 92 | 28 | 38 | 95 | 0.015 | 58.2 | 0.031 | 71.1 |
| $10^{-2}$ | 2935 | 4570 | 201 | 141 | 59 | 112 | 512 | 0.000 | 58.2 | 0.001 | 71.1 |

**5.2.2 Anomaly Detection Case Study.** We implement the ML anomaly-detection model (§3) in the control and data planes. We generate labeled packet-level traces from the NSL-KDD [36] dataset by expanding connection-level records to binned packet traces (i.e., each trace element represents a set of packets) and annotating them with their status (anomalous or benign). Flow-size distribution, mixing, and packet fields' rates of change are sampled from the original traces to create a realistic workload. We use this data to train the DNN, with an offline F1 score (a typical ML metric [157]) of 71.1.

To test data-plane ML, we route all traffic through the switch and FPGA. The switch preprocesses features using a MAT. It first uses the packet's five-tuple to index a set of stateful registers, which accumulate features across packets (e.g., the number of urgent flags). Then, it formats these features as fixed-point numbers and forwards packets to the FPGA, where the ML model marks them as anomalous or benign. The packets then return to the switch, where a postprocessing MAT interprets the ML decision.

For our baseline, the switch accumulates features and sends them as telemetry packets to the control plane for inference. If the model determines that a packet is anomalous, the respective IP is extracted and a switch rule is installed marking its packets as anomalous. Any packets that pass through before rule installation are (incorrectly) forwarded as-is. Traffic is sent at a fixed 5 Gbps while control-plane sampling varies from 100 kbps ($10^{-5}$) to 100 Mbps ($10^{-2}$).

***Taurus Responds Per-Packet.*** Table 8 shows the control-plane's millisecond-range latencies. Even at low sampling rates, the baseline has a 32 ms latency. As load grows, so do batch sizes—this increases latency, because the batch's first element must wait for the entire batch to finish. Furthermore, when approaching a per-packet system, ML is not the bottleneck. Rather, Table 8 shows that rule installation and packet collection overwhelm the system.[7]

***Taurus Sustains Full Model Accuracy.*** Most anomalous packets are undetected by the baseline, while Taurus captures over half (Table 8). We use an F1 score to evaluate accuracy, which takes into account the number of identified anomalies, missed anomalies, and benign packets incorrectly marked as anomalous [8, 153]. Taurus achieves the same F1 score as the model in isolation. However, the baseline misses packets due to the rule-installation and other processing delays, yielding a lower effective F1 score.

[7]We also explore rule installation at the switch CPU instead of the server, to eliminate the RTT between controller and switch OS. However, since the switch CPU is less powerful (8 cores at 1.6 GHz vs. an 80-core server at 2.5 GHz), the reduced RTT time is offset by additional flow-rule computation time, with 112% higher latency on average.

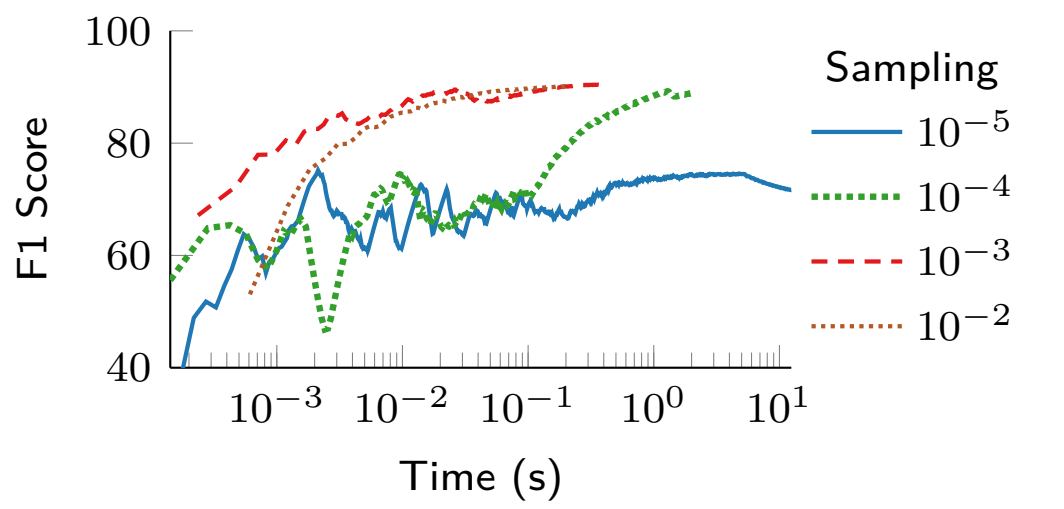

**Figure 13: Taurus's accuracy improves over time. Training with higher sampling rates converges faster.**

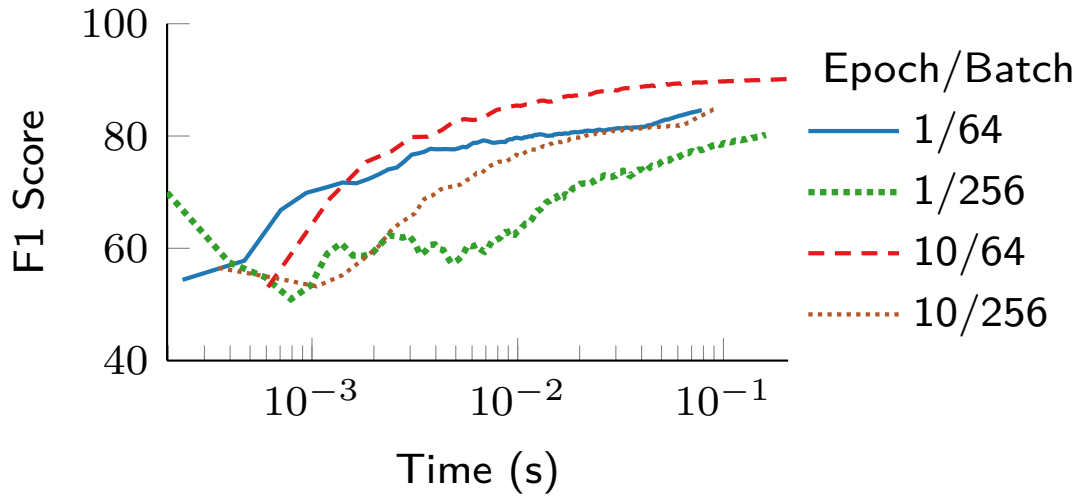

**Figure 14: Training with smaller batches and more epochs converges faster (sampling rate is $10^{-2}$).**

**5.2.3 Online Training.** Taurus's ML models can also be updated to optimize global metrics; this accounts for behavior (like downstream congestion) that cannot be observed at a single switch. We feed telemetry packets to a control-plane training application and evaluate, using flow-rule installation time as an estimate, the time needed to update data-plane model weights.

Figure 13 shows that higher sampling rates (corresponding to larger batches) converge faster (tens to hundreds of milliseconds), demonstrating that online training is possible and is aided by higher packet ingestion rates at the control plane. Moreover, Figure 14 shows that the configuration with the smallest batch size (64) and most epochs (10) results in convergence to the highest F1 score and that the added time to train is offset by faster convergence. Hence, smaller batches and larger epochs lead to less frequent but more substantial updates as opposed to frequent, less informed updates.

## 6 LIMITATIONS & FUTURE WORK

While Taurus can support per-packet ML algorithms in a more efficient manner than existing platforms, the model sizes are still limited by the available hardware resources. To fit larger models, we must look into model compression techniques. Furthermore, additional research is needed to provide provably-correct model assertions and faster training times.

***Shrinking Models.*** A major Taurus application will be network control and coordination. Neural networks can solve a variety of control problems [10, 149, 156] and are getting smaller. For example, structured control nets [149] for non-linear control perform almost as well as 512-neuron DNNs using as few as four neurons per layer. With such small networks, Taurus can run multiple models simultaneously (e.g., one model for intrusion detection and another for traffic optimization). In addition, techniques like quantization, pruning, and distillation can further reduce a models' size [7, 69, 81, 158].

***Correctness.*** Some applications, like routing, have definitive correct and incorrect answers and are hard to guarantee without safeguards. Instead, ML is best suited for applications that are inherently heuristic, such as congestion control [162, 169], load balancing [4, 85], and anomaly detection [106, 153]; their decisions impact only networks' performance and security, not their core packet-forwarding behavior. When an ML decision could impact network correctness, postprocessing rules can ensure that the final decision is bounded.

***Training Speed.*** Finally, the training reaction time would likely be higher for a larger datacenter network. This means that instantaneous network problems (like load on a specific link) cannot be handled via training—the event would be over before training completes. Instead, training gradually learns what causes these problems and how to avoid them. To detect and react to specific events, lower-latency techniques, like INT, can instead provide network status to the model as features.

## 7 RELATED WORK

***Architectures for ML.*** Field programmable gate arrays (FPGAs) are the most widely-available reconfigurable architectures, used as both custom accelerators [140] and prototyping tools (e.g., the NetFPGA [104, 113]). However, FPGAs' on-chip interconnects consume up to 70% of the total chip power [23], and their variable, slow clock frequencies complicate interfacing and operating at network switch speeds (multi-terabits per second). CGRAs are optimized for ML arithmetic and typically have a fast, fixed clock frequency that allows seamless integration between the MapReduce block and MATs in Taurus [32, 50, 58, 105, 107, 143]. Other architectures, like Eyeriss [26], Brainwave [49], and EIE [68], achieve high efficiency by focusing on specific algorithms. These could be used for in-switch ML, but are too rigid: if a specific accelerator were standardized, networks would be unable to benefit from future ML research due to the lack of a flexible abstraction (like MapReduce).

***Data Planes for ML.*** Inside a datacenter core, programmable MAT-based data-plane switches (like Barefoot Tofino [114, 115]), today, allow networks to easily support and perform tasks (such as heavy-hitter detection [148], load balancing [85], security [93, 175], fast rerouting [74], and scheduling [147]), previously implemented using end-host servers, middleboxes, or fixed-function switches. Recent efforts (N2Net [144] and IIsy [168]) are now looking into running more complicated machine-learning algorithms on these switches. However, MATs alone are ill-suited to support ML algorithms: for data-plane ML to become ubiquitous, we need more efficient hardware to minimize unnecessary resource usage (i.e., area and power). We believe Taurus is a first step in that direction.

At the datacenter end-hosts, modern SmartNICs (and network accelerators) provide higher computational capacity for servers to offload not only packet-processing logic but also application-specific logic closer to the network interface for lower latency and higher throughput. Products like Intel's Infrastructure Units (IPUs) [80], NVIDIA's Data Processing Units (DPUs) [120], and FPGA-based SmartNICs [44, 164] all aim at accelerating a variety of end-host workloads (e.g., hypervisors, virtual switching, batch-based ML training and inference, storage). However, these SmartNICs are a poor fit for per-packet operations that need to happen at a network-wide scale in a data center. Packets would still have to visit the server (and NIC) from the switch, resulting in an RTT worth of delay between when a packet is seen and when the corresponding decision is applied at the switch. Middleboxes experience similar issues on top of their fixed-function hardware implementation, which limits flexibility and prompts frequent upgrades [141].

***ML for Networking.*** Many networking applications can benefit from ML. For example, learned algorithms for congestion control [162, 169] have been shown to outperform their human-designed counterparts [37, 66, 171]. In addition, Boutaba et al. [17] identify ML use cases for network tasks such as traffic classification [40, 41], traffic prediction [24, 27], active queue management [151, 179], and security [124]. All of these applications could immediately be deployed using Taurus.

***Networking for ML.*** Specialized networks can also accelerate ML algorithms themselves. With minor enhancements to modern data-plane hardware, switches can aggregate gradients in-network, accelerating training by up to 300% [51, 92, 137]. Gaia, a system for distributed ML [76], also accounts for wide-area network bandwidth and regulates the movements of gradients during the training process. While Taurus is not explicitly designed to accelerate distributed training, MapReduce can aggregate numeric weights, contained in packets, more efficiently than MATs.

## 8 CONCLUSION

Currently, datacenter networks are partitioned into a line-rate, per-packet data plane and a slower control plane, where the control plane runs complex, data-driven management policies to configure the data plane. This approach is too slow because transiting the control plane adds unavoidable latency, so the control plane cannot react quickly enough to network events. Taurus brings complex decision-making to the data plane, bringing the reaction time down to per-packet and increasing the specificity of managemnet and control policies. We demonstrate that Taurus operates at line rate and adds minimal overhead to a programmable switch pipeline (RMT)—3.8% more area and 122 ns average latency—while accelerating several recently-proposed ML networking benchmarks.

Looking forward, to realize a self-driving network, hardware must be deployed before large-scale training can begin. We believe, Taurus gives a foothold for in-network ML with hardware that can be installed in next-generation data planes—to improve performance and security.

# A ARTIFACT APPENDIX

## A.1 Abstract

The artifact contains the source code for the proposed Taurus's MapReduce block (§3.3) and the anomaly-detection (AD) application used in the end-to-end performance evaluation (§5.2). The MapReduce block is integrated with the Xilinx OpenNIC Shell [166] and is programmed via a high-level DSL, called Spatial [88]. The AD application runs on the end-to-end testbed and consists of various components: P4 programs, Python scripts, ONOS application, and the Spatial code (§5.2.2).

## A.2 Scope

The artifact provides the two new contributions: MapReduce block and the anomaly-detection code. The goal is to provide the two missing components needed to implement the complete testbed (§5.2) for running the end-to-end performance evaluation. The remaining components are a collection of existing propriety and specialized hardware (Barefoot Tofino Switch [114], Xilinx Alveo Board [164]) and software codebases (P4 [14], ONOS [46], Spatial [88], and more), which are readily available for purchase and download (please see the dependencies, below).[8]

## A.3 Contents

- *MapReduce Block in FPGA:* This repository contains the source code and instructions for building an FPGA-based implementation of the MapReduce block. The detailed documentation for running the MapReduce block on FPGA is provided here: https://gitlab.com/dataplane-ai/taurus/mapreduce.
- *Anomaly-Detection Application:* In this repository, we share the source code (P4, Python, ONOS, Spatial) for the anomaly-detection application (AD). We also provide details on what is needed to replicate the end-to-end testbed used for evaluating the AD application. The description of the AD application is located here: https://gitlab.com/dataplane-ai/taurus/applications/anomaly-detection-asplos22.

## A.4 Hosting

The source code is publicly available on GitLab[9] and figshare.[10]

## A.5 Dependencies

The artifact relies on the following third-party hardware and software tools:

- EdgeCore Wedge 100BF-32X Switch
- Xilinx Alveo U250 FPGA board
- Intel P4 Studio
- Xilinx Vivado Design Tool
- Xilinx OpenNIC Shell
- Spatial DSL
- ONF Stratum OS
- ONF Open Network Operating System (ONOS)
- MoonGen Traffic Generator

[8]We expect users to buy the required hardware and gain individual access to the necessary tools and licenses and, therefore, do not expand on setting up these components in the artifact.

[9]https://gitlab.com/dataplane-ai/taurus

[10]https://doi.org/10.6084/m9.figshare.17097524

# REFERENCES

[1] Martín Abadi, Paul Barham, Jianmin Chen, Zhifeng Chen, Andy Davis, Jeffrey Dean, Matthieu Devin, Sanjay Ghemawat, Geoffrey Irving, Michael Isard, et al. 2016. Tensorflow: A System For Large-Scale Machine Learning. In *USENIX OSDI '16*.

[2] Preeti Aggarwal and Sudhir Kumar Sharma. 2015. Analysis of KDD Dataset Attributes-Class Wise For Intrusion Detection. *Computer Science* 57 (2015), 842–851. https://doi.org/10.1016/j.procs.2015.07.490

[3] Anurag Agrawal and Changhoon Kim. 2020. Intel Tofino2 – A 12.9Tbps P4-Programmable Ethernet Switch. In *Hot Chips '20*.

[4] Mohammad Alizadeh, Tom Edsall, Sarang Dharmapurikar, Ramanan Vaidyanathan, Kevin Chu, Andy Fingerhut, Vinh The Lam, Francis Matus, Rong Pan, Navindra Yadav, and George Varghese. 2014. CONGA: Distributed Congestion-aware Load Balancing for Datacenters. In *ACM SIGCOMM '14*. https://doi.org/10.1145/2619239.2626316

[5] Mohammad Alizadeh, Albert Greenberg, David A Maltz, Jitendra Padhye, Parveen Patel, Balaji Prabhakar, Sudipta Sengupta, and Murari Sridharan. 2010. Data Center TCP (DCTCP). In *ACM SIGCOMM '10*. https://doi.org/10.1145/1851182.1851192

[6] Tom Auld, Andrew W Moore, and Stephen F Gull. 2007. Bayesian Neural Networks For Internet Traffic Classification. *IEEE Transactions on Neural Networks '07* 18, 1 (2007), 223–239. https://doi.org/10.1109/TNN.2006.883010

[7] Jimmy Ba and Rich Caruana. 2014. Do Deep Nets Really Need to be Deep?. In *NeurIPS '14*.

[8] Jarrod Bakker, Bryan Ng, Winston KG Seah, and Adrian Pekar. 2019. Traffic Classification with Machine Learning in a Live Network. In *2019 IFIP/IEEE Symposium on Integrated Network and Service Management (IM) '19*.

[9] Rajeev Balasubramonian, Andrew B Kahng, Naveen Muralimanohar, Ali Shafiee, and Vaishnav Srinivas. 2017. CACTI 7: New tools for interconnect exploration in innovative off-chip memories. *ACM Transactions on Architecture and Code Optimization (TACO '17)* 14, 2 (2017), 1–25. https://doi.org/10.1145/3085572

[10] Marc G Bellemare, Yavar Naddaf, Joel Veness, and Michael Bowling. 2013. The Arcade Learning Environment: An Evaluation Platform for General Agents. *Journal of Artificial Intelligence Research (JAIR)* 47 (2013), 253–279.

[11] Laurent Bernaille, Renata Teixeira, Ismael Akodkenou, Augustin Soule, and Kave Salamatian. 2006. Traffic Classification on the Fly. *ACM SIGCOMM Computer Communication Review (CCR)* 36, 2 (2006), 23–26. https://doi.org/10.1145/1129582.1129589

[12] Kirti Bhanushali and W Rhett Davis. 2015. FreePDK15: An open-source predictive process design kit for 15nm FinFET technology. In *Proceedings of the 2015 Symposium on International Symposium on Physical Design*. 165–170. https://doi.org/10.1145/2717764.2717782

[13] Pat Bosshart, Dan Daly, Glen Gibb, Martin Izzard, Nick McKeown, Jennifer Rexford, Cole Schlesinger, Dan Talayco, Amin Vahdat, George Varghese, et al. 2014. P4: Programming protocol-independent packet processors. *ACM SIGCOMM Computer Communication Review* 44, 3 (2014), 87–95. https://doi.org/10.1145/2656877.2656890

[14] Pat Bosshart, Dan Daly, Glen Gibb, Martin Izzard, Nick McKeown, Jennifer Rexford, Cole Schlesinger, Dan Talayco, Amin Vahdat, George Varghese, et al. 2014. P4: Programming Protocol-Independent Packet Processors. *ACM SIGCOMM Computer Communication Review (CCR)* 44, 3 (2014), 87–95. https://doi.org/10.1145/2656877.2656890

[15] Pat Bosshart, Glen Gibb, Hun-Seok Kim, George Varghese, Nick McKeown, Martin Izzard, Fernando Mujica, and Mark Horowitz. 2013. Forwarding Metamorphosis: Fast Programmable Match-Action Processing in Hardware for SDN. In *ACM SIGCOMM '13*. https://doi.org/10.1145/2534169.2486011

[16] Leon Bottou. 2010. Feature Engineering. https://www.cs.princeton.edu/ courses/archive/spring10/cos424/slides/18-feat.pdf. Accessed on 08/12/2021.

[17] Raouf Boutaba, Mohammad A Salahuddin, Noura Limam, Sara Ayoubi, Nashid Shahriar, Felipe Estrada-Solano, and Oscar M Caicedo. 2018. A Comprehensive Survey on Machine Learning for Networking: Evolution, Applications and Research Opportunities. *Journal of Internet Services and Applications (JISA)* 9, 1 (2018), 16. https://doi.org/10.1186/s13174-018-0087-2

[18] Andrey Brito, Andre Martin, Thomas Knauth, Stephan Creutz, Diogo Becker, Stefan Weigert, and Christof Fetzer. 2011. Scalable and Low-Latency Data

Processing with Stream MapReduce. In *IEEE CLOUDCOM '11*. https://doi.org/10.1109/CloudCom.2011.17

[19] Broadcom. [n.d.]. Tomahawk/BCM56960 Series. https://www.broadcom.com/products/ethernet-connectivity/switching/strataxgs/bcm56960-series. Accessed on 08/12/2021.

[20] Kevin J Brown, HyoukJoong Lee, Tiark Romp, Arvind K Sujeeth, Christopher De Sa, Christopher Aberger, and Kunle Olukotun. 2016. Have Abstraction and Eat Performance, too: Optimized Heterogeneous Computing with Parallel Patterns. In *IEEE/ACM CGO '16*. https://doi.org/10.1145/2854038.2854042

[21] Kevin J Brown, Arvind K Sujeeth, Hyouk Joong Lee, Tiark Rompf, Hassan Chafi, Martin Odersky, and Kunle Olukotun. 2011. A Heterogeneous Parallel Framework for Domain-Specific Languages. In *IEEE PACT '11*. https://doi.org/10.1109/PACT.2011.15

[22] Douglas Ronald Burdick, Amol Ghoting, Rajasekar Krishnamurthy, Edwin Peter Dawson Pednault, Berthold Reinwald, Vikas Sindhwani, Shirish Tatikonda, Yuanyuan Tian, and Shivakumar Vaithyanathan. 2013. Systems and methods for processing machine learning algorithms in a MapReduce environment. US Patent 8,612,368.

[23] Benton Highsmith Calhoun, Joseph F Ryan, Sudhanshu Khanna, Mateja Putic, and John Lach. 2010. Flexible Circuits and Architectures for Ultralow Power. *Proc. IEEE* 98, 2 (2010), 267–282.

[24] Samira Chabaa, Abdelouhab Zeroual, and Jilali Antari. 2010. Identification and Prediction of Internet Traffic Using Artificial Neural Networks. *Journal of Intelligent Learning Systems and Applications (JILSA)* 2, 03 (2010), 147. https://doi.org/10.4236/jilsa.2010.23018

[25] Huan Chen and Theophilus Benson. 2017. The Case for Making Tight Control Plane Latency Guarantees in SDN Switches. In *ACM SOSR '17*. https://doi.org/10.1145/3050220.3050237

[26] Yu-Hsin Chen, Tushar Krishna, Joel S Emer, and Vivienne Sze. 2016. Eyeriss: An Energy-Efficient Reconfigurable Accelerator for Deep Convolutional Neural Networks. *IEEE Journal of Solid-State Circuits* 52, 1 (2016), 127–138. https://doi.org/10.1109/JSSC.2016.2616357

[27] Zhitang Chen, Jiayao Wen, and Yanhui Geng. 2016. Predicting Future Traffic Using Hidden Markov Models. In *IEEE ICNP '16*.

[28] Cheng-Tao Chu, Sang K Kim, Yi-An Lin, YuanYuan Yu, Gary Bradski, Kunle Olukotun, and Andrew Y Ng. 2007. Map-Reduce for Machine Learning on Multicore. In *NeurIPS '07*. 281–288.

[29] Cisco Systems, Inc. [n.d.]. Cisco Meraki (MX450): Powerful Security and SD-WAN for the Branch & Campus. https://meraki.cisco.com/products/appliances/mx450. Accessed on 08/12/2021.

[30] Graham Cormode and Shan Muthukrishnan. 2005. An Improved Data Stream Summary: The Count-Min Sketch and its Applications. *Journal of Algorithms* 55, 1 (2005), 58–75. https://doi.org/10.1016/j.jalgor.2003.12.001

[31] Paul Covington, Jay Adams, and Emre Sargin. 2016. Deep Neural Networks for Youtube Recommendations. In *ACM RecSys '16*. https://doi.org/10.1145/2959100.2959190

[32] Darren C Cronquist, Chris Fisher, Miguel Figueroa, Paul Franklin, and Carl Ebeling. 1999. Architecture Design of Reconfigurable Pipelined Datapaths. In *IEEE ARVLSI '99*.

[33] Influx Data. [n.d.]. InfluxDB. https://www.influxdata.com/products/influxdb-overview/. Accessed on 08/12/2021.

[34] Christopher De Sa, Megan Leszczynski, Jian Zhang, Alana Marzoev, Christopher R Aberger, Kunle Olukotun, and Christopher Ré. 2018. High-Accuracy Low-Precision Training. *arXiv preprint arXiv:1803.03383* (2018).

[35] Dell EMC. [n.d.]. Data Center Switching Quick Reference Guide. https://i.dell.com/sites/doccontent/shared-content/data-sheets/en/Documents/Dell-Networking-Data-Center-Quick-Reference-Guide.pdf. Accessed on 08/12/2021.

[36] L Dhanabal and SP Shantharajah. 2015. A Study on NSL-KDD Dataset for Intrusion Detection System Based on Classification Algorithms. *International Journal of Advanced Research in Computer and Communication Engineering (IJARCCE)* 4, 6 (2015), 446–452. https://doi.org/10.17148/IJARCCE.2015.4696

[37] Mo Dong, Qingxi Li, Doron Zarchy, P Brighten Godfrey, and Michael Schapira. 2015. PCC: Re-architecting Congestion Control for Consistent High Performance. In *USENIX NSDI '15*.

[38] DPDK. [n.d.]. DPDK. https://www.dpdk.org/. Accessed on 08/12/2021.

[39] Paul Emmerich, Sebastian Gallenmüller, Daniel Raumer, Florian Wohlfart, and Georg Carle. 2015. Moongen: A Scriptable High-Speed Packet Generator. In *ACM IMC '15*.

[40] Jeffrey Erman, Martin Arlitt, and Anirban Mahanti. 2006. Traffic Classification Using Clustering Algorithms. In *ACM MineNet '06*. https://doi.org/10.1145/1162678.1162679

[41] Jeffrey Erman, Anirban Mahanti, Martin Arlitt, and Carey Williamson. 2007. Identifying and Discriminating Between Web and Peer-to-Peer Traffic in the Network Core. In *WWW '07*. https://doi.org/10.1145/1242572.1242692

[42] Alice Este, Francesco Gringoli, and Luca Salgarelli. 2009. Support Vector Machines For TCP Traffic Classification. *Computer Networks* 53, 14 (2009), 2476–2490. https://doi.org/10.1016/j.comnet.2009.05.003

[43] Nick Feamster and Jennifer Rexford. 2018. Why (and How) Networks Should Run Themselves. In *ACM ANRW '18*.

[44] Daniel Firestone, Andrew Putnam, Sambhrama Mundkur, Derek Chiou, Alireza Dabagh, Mike Andrewartha, Hari Angepat, Vivek Bhanu, Adrian Caulfield, Eric Chung, et al. 2018. Azure Accelerated Networking: SmartNICs in the Public Cloud. In *USENIX NSDI '18*.

[45] Sally Floyd and Van Jacobson. 1993. Random early detection gateways for congestion avoidance. *IEEE/ACM Transactions on networking* 1, 4 (1993), 397–413.

[46] Open Networking Foundation. [n.d.]. ONOS: Open Network Operating System. https://www.opennetworking.org/onos/. Accessed on 08/12/2021.

[47] Open Networking Foundation. [n.d.]. ONOS: Single Bench Flow Latency Test. https://wiki.onosproject.org/display/ONOS/2.2%3A+Experiment+I+-+Single+Bench+Flow+Latency+Test. Accessed on 08/12/2021.

[48] Open Networking Foundation. [n.d.]. Stratum OS. https://www.opennetworking.org/stratum/. Accessed on 08/12/2021.

[49] Jeremy Fowers, Kalin Ovtcharov, Michael Papamichael, Todd Massengill, Ming Liu, Daniel Lo, Shlomi Alkalay, Michael Haselman, Logan Adams, Mahdi Ghandi, et al. 2018. A Configurable Cloud-Scale DNN Processor for Real-Time AI. In *IEEE ISCA '18*. https://doi.org/10.1109/ISCA.2018.00012

[50] Mingyu Gao and Christos Kozyrakis. 2016. HRL: Efficient and Flexible Reconfigurable Logic for Near-Data Processing. In *IEEE HPCA '16*.

[51] Nadeen Gebara, Manya Ghobadi, and Paolo Costa. 2021. In-network Aggregation for Shared Machine Learning Clusters. In *MLSys '21*.

[52] Yilong Geng, Shiyu Liu, Feiran Wang, Zi Yin, Balaji Prabhakar, and Mendel Rosenblum. 2017. Self-Programming Networks: Architecture and Algorithms. In *IEEE Allerton '17*.

[53] Yilong Geng, Shiyu Liu, Zi Yin, Ashish Naik, Balaji Prabhakar, Mendel Rosenblum, and Amin Vahdat. 2019. SIMON: A Simple and Scalable Method for Sensing, Inference and Measurement in Data Center Networks. In *USENIX NSDI '19*.

[54] Amol Ghoting, Prabhanjan Kambadur, Edwin Pednault, and Ramakrishnan Kannan. 2011. NIMBLE: A Toolkit for the Implementation of Parallel Data Mining and Machine Learning Algorithms on MapReduce. In *ACM SIGKDD KDD '11*. https://doi.org/10.1145/2020408.2020464

[55] Amol Ghoting, Rajasekar Krishnamurthy, Edwin Pednault, Berthold Reinwald, Vikas Sindhwani, Shirish Tatikonda, Yuanyuan Tian, and Shivakumar Vaithyanathan. 2011. SystemML: Declarative Machine Learning on MapReduce. In *IEEE ICDE '11*.

[56] Glen Gibb, George Varghese, Mark Horowitz, and Nick McKeown. 2013. Design principles for packet parsers. In *ACM/IEE ANCS '13*. https://doi.org/10.1109/ANCS.2013.6665172

[57] Dan Gillick, Arlo Faria, and John DeNero. 2006. MapReduce: Distributed Computing for Machine Learning. *Berkley, Dec* 18 (2006).

[58] Seth Copen Goldstein, Herman Schmit, Mihai Budiu, Srihari Cadambi, Matthew Moe, and R Reed Taylor. 2000. PipeRench: A Reconfigurable Architecture and Compiler. *Computer* 33, 4 (2000), 70–77.

[59] Ian Goodfellow, Yoshua Bengio, Aaron Courville, and Yoshua Bengio. 2016. *Deep Learning*. Vol. 1. MIT Press, Cambridge.

[60] Google. [n.d.]. A look inside Google's Data Center Networks. https://cloudplatform.googleblog.com/2015/06/A-Look-Inside-Googles-Data-Center-Networks.html. Accessed on 08/12/2021.

[61] Google. 2021. Tensorflow Lite. https://www.tensorflow.org/tflite.

[62] Albert Greenberg, James R. Hamilton, Navendu Jain, Srikanth Kandula, Changhoon Kim, Parantap Lahiri, David A. Maltz, Parveen Patel, and Sudipta Sengupta. 2009. VL2: A Scalable and Flexible Data Center Network. In *ACM SIGCOMM '09*. https://doi.org/10.1145/1592568.1592576

[63] P4.org Archictecture Working Group. 2018. P4-16 Portable Switch Architecture. https://p4.org/p4-spec/docs/PSA-v1.1.0.pdf.

[64] Antonio Gulli and Sujit Pal. 2017. *Deep learning with Keras*. Packt Publishing Ltd.

[65] Vladimir Gurevich. 2018. Programmable Data Plane at Terabit Speeds. https://p4.org/assets/p4_d2_2017_programmable_data_plane_at_terabit_speeds.pdf. Accessed on 08/12/2021.

[66] Sangtae Ha, Injong Rhee, and Lisong Xu. 2008. CUBIC: A New TCP-Friendly High-Speed TCP Variant. *ACM SIGOPS Operating Systems Review* 42, 5 (2008), 64–74. https://doi.org/10.1145/1400097.1400105

[67] Song Han, Junlong Kang, Huizi Mao, Yiming Hu, Xin Li, Yubin Li, Dongliang Xie, Hong Luo, Song Yao, Yu Wang, et al. 2017. ESE: Efficient Speech Recognition Engine with Sparse LSTM on FPGA. In *ACM/SIGDA FPGA '17*. https://doi.org/10.1145/3020078.3021745

[68] Song Han, Xingyu Liu, Huizi Mao, Jing Pu, Ardavan Pedram, Mark A Horowitz, and William J Dally. 2016. EIE: Efficient Inference Engine on Compressed Deep Neural Network. In *ACM/IEEE ISCA '16*. https://doi.org/10.1145/3007787.3001163

[69] Song Han, Jeff Pool, John Tran, and William Dally. 2015. Learning Both Weights and Connections for Efficient Neural Network. In *NeurIPS '15*.

[70] B Hariri and N Sadati. 2007. NN-RED: an AQM mechanism based on neural networks. *Electronics Letters* 43, 19 (2007), 1053–1055. https://doi.org/10.1049/el:

20071791
[71] Robert Harper, David MacQueen, and Robin Milner. 1986. *Standard ML*. Department of Computer Science, University of Edinburgh.
[72] Marti A. Hearst, Susan T Dumais, Edgar Osuna, John Platt, and Bernhard Scholkopf. 1998. Support Vector Machines (SVMs). *IEEE Intelligent Systems and their Applications* 13, 4 (1998), 18–28.
[73] Sepp Hochreiter and Jürgen Schmidhuber. 1997. Long Short-Term Memory (LSTM). *Neural Computation* 9, 8 (1997), 1735–1780.
[74] Thomas Holterbach, Edgar Costa Molero, Maria Apostolaki, Alberto Dainotti, Stefano Vissicchio, and Laurent Vanbever. 2019. Blink: Fast Connectivity Recovery Entirely in the Data Plane. In *USENIX NSDI '19*.
[75] Kurt Hornik. 1991. Approximation capabilities of multilayer feedforward networks. *Neural networks* 4, 2 (1991), 251–257. https://doi.org/10.1016/0893-6080(91)90009-T
[76] Kevin Hsieh, Aaron Harlap, Nandita Vijaykumar, Dimitris Konomis, Gregory R Ganger, Phillip B Gibbons, and Onur Mutlu. 2017. Gaia: Geo-Distributed Machine Learning Approaching LAN Speeds. In *USENIX NSDI '17*.
[77] Peng Huang, Chuanxiong Guo, Lidong Zhou, Jacob R Lorch, Yingnong Dang, Murali Chintalapati, and Randolph Yao. 2017. Gray Failure: The Achilles' Heel of Cloud-Scale Systems. In *ACM HotOS '17*. https://doi.org/10.1145/3102980.3103005
[78] Intel. [n.d.]. Intel Deep Insight Network Analytics Software. https://www.intel.com/content/www/us/en/products/network-io/programmable-ethernet-switch/network-analytics/deep-insight.html. Accessed on 08/12/2021.
[79] Intel. [n.d.]. Intel® Ethernet Network Adapter X710-DA2 for OCP 3.0. https://ark.intel.com/content/www/us/en/ark/products/184822/intel-ethernet-network-adapter-x710-da4-for-ocp-3-0.html. Accessed on 08/12/2021.
[80] Intel. 2021. Intel Infrastructure Processing Unit (Intel IPU) and SmartNICs. https://www.intel.com/content/www/us/en/products/network-io/smartnic.html.
[81] Benoit Jacob, Skirmantas Kligys, Bo Chen, Menglong Zhu, Matthew Tang, Andrew Howard, Hartwig Adam, and Dmitry Kalenichenko. 2018. Quantization and Training of Neural Networks for Efficient Integer-Arithmetic-Only Inference. In *IEEE CVPR '18*.
[82] Nathan Jay, Noga Rotman, Brighten Godfrey, Michael Schapira, and Aviv Tamar. 2019. A deep reinforcement learning perspective on internet congestion control. In *ICML '19*.
[83] Eun Young Jeong, Shinae Woo, Muhammad Jamshed, Haewon Jeong, Sunghwan Ihm, Dongsu Han, and KyoungSoo Park. 2014. MTCP: A Highly Scalable User-Level TCP Stack for Multicore Systems. In *USENIX NSDI '14*.
[84] Norman P Jouppi, Cliff Young, Nishant Patil, David Patterson, Gaurav Agrawal, Raminder Bajwa, Sarah Bates, Suresh Bhatia, Nan Boden, Al Borchers, et al. 2017. In-Datacenter Performance Analysis of a Tensor Processing Unit. In *IEEE ISCA '17*. https://doi.org/10.1145/3079856.3080246
[85] Naga Katta, Mukesh Hira, Changhoon Kim, Anirudh Sivaraman, and Jennifer Rexford. 2016. HULA: Scalable Load Balancing Using Programmable Data Planes. In *ACM SOSR '16*. https://doi.org/10.1145/2890955.2890968
[86] Changhoon Kim. [n.d.]. Programming The Network Data Plane: What, How, and Why? https://conferences.sigcomm.org/ events/apnet2017/slides/chang.pdf. Accessed on 08/12/2021.
[87] Changhoon Kim, Anirudh Sivaraman, Naga Katta, Antonin Bas, Advait Dixit, and Lawrence J Wobker. 2015. In-Band Network Telemetry via Programmable Dataplanes. In *ACM SIGCOMM '15 (Demo)*.
[88] David Koeplinger, Matthew Feldman, Raghu Prabhakar, Yaqi Zhang, Stefan Hadjis, Ruben Fiszel, Tian Zhao, Luigi Nardi, Ardavan Pedram, Christos Kozyrakis, and Kunle Olukotun. 2018. Spatial: A Language and Compiler for Application Accelerators. In *ACM/SIGPLAN PLDI '18*. https://doi.org/10.1145/3192366.3192379
[89] Vibhore Kumar, Henrique Andrade, Buğra Gedik, and Kun-Lung Wu. 2010. DEDUCE: At the Intersection of MapReduce and Stream Processing. In *ACM EDBT '10*. https://doi.org/10.1145/1739041.1739120
[90] Maciej Kuźniar, Peter Perešíni, and Dejan Kostić. 2015. What You Need to Know About SDN Flow Tables. In *PAM '15*. Springer.
[91] Adam Langley, Alistair Riddoch, Alyssa Wilk, Antonio Vicente, Charles Krasic, Dan Zhang, Fan Yang, Fedor Kouranov, Ian Swett, Janardhan Iyengar, et al. 2017. The QUIC Transport Protocol: Design and Internet-Scale Deployment. In *ACM SIGCOMM '17*. https://doi.org/10.1145/3098822.3098842
[92] ChonLam Lao, Yanfang Le, Kshiteej Mahajan, Yixi Chen, Wenfei Wu, Aditya Akella, and Michael M Swift. 2021. ATP: In-network Aggregation for Multi-tenant Learning. In *NSDI '21*.
[93] Ângelo Cardoso Lapolli, Jonatas Adilson Marques, and Luciano Paschoal Gaspary. 2019. Offloading real-time ddos attack detection to programmable data planes. In *2019 IFIP/IEEE Symposium on Integrated Network and Service Management (IM) '19*. IEEE, 19–27.
[94] Yann LeCun, LD Jackel, Leon Bottou, A Brunot, Corinna Cortes, JS Denker, Harris Drucker, I Guyon, UA Muller, Eduard Sackinger, P Simard, and V Vapnik. 1995. Comparison of Learning Algorithms for Handwritten Digit Recognition. In *ICANN '95*.
[95] Guanyu Li, Menghao Zhang, Shicheng Wang, Chang Liu, Mingwei Xu, Ang Chen, Hongxin Hu, Guofei Gu, Qi Li, and Jianping Wu. 2021. Enabling Performant, Flexible and Cost-Efficient DDoS Defense With Programmable Switches. *IEEE/ACM Transactions on Networking '21* (2021). https://doi.org/10.1109/TNET.2021.3062621
[96] Ruey-Hsia Li and Geneva G. Belford. 2002. Instability of Decision Tree Classification Algorithms. In *ACM SIGKDD '02*. https://doi.org/10.1145/775047.775131
[97] Youjie Li, Iou-Jen Liu, Yifan Yuan, Deming Chen, Alexander Schwing, and Jian Huang. 2019. Accelerating Distributed Reinforcement Learning with In-Switch Computing. In *ACM/IEEE ISCA '19*. https://doi.org/10.1145/3307650.3322259
[98] Yuliang Li, Rui Miao, Hongqiang Harry Liu, Yan Zhuang, Fei Feng, Lingbo Tang, Zheng Cao, Ming Zhang, Frank Kelly, Mohammad Alizadeh, and Minlan Yu. 2019. HPCC: High Precision Congestion Control. In *ACM SIGCOMM '19*. https://doi.org/10.1145/3341302.3342085
[99] Eric Liang, Hang Zhu, Xin Jin, and Ion Stoica. 2019. Neural Packet Classification. In *ACM SIGCOMM '19*. https://doi.org/10.1145/3341302.3342221
[100] Darryl Lin, Sachin Talathi, and Sreekanth Annapureddy. 2016. Fixed Point Quantization of Deep Convolutional Networks. In *ICML '16*.
[101] Yingqiu Liu, Wei Li, and Yun-Chun Li. 2007. Network Traffic Classification Using K-Means Clustering. In *IEEE IMSCCS '07*. https://doi.org/10.1109/IMSCCS.2007.52
[102] Zaoxing Liu, Hun Namkung, Georgios Nikolaidis, Jeongkeun Lee, Changhoon Kim, Xin Jin, Vladimir Braverman, Minlan Yu, and Vyas Sekar. 2021. Jaqen: A High-Performance Switch-Native Approach for Detecting and Mitigating Volumetric DDoS Attacks with Programmable Switches. In *USENIX Security '21*.
[103] Loadbalancer.org. [n.d.]. Hardware ADC. https://www.loadbalancer.org/products/hardware/. Accessed on 08/12/2021.
[104] John W Lockwood, Nick McKeown, Greg Watson, Glen Gibb, Paul Hartke, Jad Naous, Ramanan Raghuraman, and Jianying Luo. 2007. NetFPGA: An Open Platform for Gigabit-Rate Network Switching and Routing. In *IEEE MSE '07*.
[105] Alan Marshall, Tony Stansfield, Igor Kostarnov, Jean Vuillemin, and Brad Hutchings. 1999. A Reconfigurable Arithmetic Array for Multimedia Applications. In *IEEE FPGA '99*. https://doi.org/10.1145/296399.296444
[106] Tahir Mehmood and Helmi B Md Rais. 2015. SVM for Network Anomaly Detection using ACO Feature Subset. In *IEEE iSMSC '15*.
[107] Bingfeng Mei, Serge Vernalde, Diederik Verkest, Hugo De Man, and Rudy Lauwereins. 2002. DRESC: A Retargetable Compiler for Coarse-Grained Reconfigurable Architectures. In *IEEE FPT '02*.
[108] Albert Mestres, Alberto Rodriguez-Natal, Josep Carner, Pere Barlet-Ros, Eduard Alarcón, Marc Solé, Victor Muntés-Mulero, David Meyer, Sharon Barkai, Mike J Hibbett, et al. 2017. Knowledge-Defined Networking. *ACM SIGCOMM Computer Communication Review (CCR)* 47, 3 (2017), 2–10. https://doi.org/10.1145/3138808.3138810
[109] Yaron Minsky, Anil Madhavapeddy, and Jason Hickey. 2013. *Real World OCaml: Functional Programming for the Masses*. O'Reilly Media, Inc.
[110] Andrew W Moore and Denis Zuev. 2005. Internet Traffic Classification using Bayesian Analysis Techniques. In *ACM SIGMETRICS '05*. https://doi.org/10.1145/1064212.1064220
[111] Manya Ghobadi Nadeen Gebara, Paolo Costa. 2021. In-Network Aggregation for Shared Machine Learning Clusters. In *MlSys*.
[112] Vinod Nair and Geoffrey E Hinton. 2010. Rectified Linear Units Improve Restricted Boltzmann Machines. In *ICML '10*.
[113] NetFPGA. [n.d.]. NetFPGA: A Line-rate, Flexible, and Open Platform for Research and Classroom Experimentation. https://netfpga.org/. Accessed on 08/12/2021.
[114] Barefoot Networks. [n.d.]. Barefoot Tofino. https://www.intel.com/content/www/us/en/products/ network-io/programmable-ethernet-switch/tofino-series.html. Accessed on 08/12/2021.
[115] Barefoot Networks. [n.d.]. Barefoot Tofino 2. https://www.intel.com/content/www/us/en/ products/network-io/programmable-ethernet-switch/tofino-2-series.html. Accessed on 08/12/2021.
[116] Edgecore Networks. [n.d.]. WEDGE 100BF-32X: 100GBE Data Center Switch. https://www.edge-core.com/productsInfo.php?cls=1&cls2=5& cls3=181&id=335. Accessed on 08/12/2021.
[117] Rolf Neugebauer, Gianni Antichi, José Fernando Zazo, Yury Audzevich, Sergio López-Buedo, and Andrew W. Moore. 2018. Understanding PCIe Performance for End Host Networking. In *ACM SIGCOMM '18*. https://doi.org/10.1145/3230543.3230560
[118] Radhika Niranjan Mysore, Andreas Pamboris, Nathan Farrington, Nelson Huang, Pardis Miri, Sivasankar Radhakrishnan, Vikram Subramanya, and Amin Vahdat. 2009. PortLand: A Scalable Fault-tolerant Layer 2 Data Center Network Fabric. In *ACM SIGCOMM '09*. https://doi.org/10.1145/1592568.1592575
[119] NVIDIA. [n.d.]. Tesla T4. https://www.nvidia.com/en-us/data-center/tesla-t4/. Accessed on 08/12/2021.
[120] NVidia. 2021. NVIDIA BLUEFIELD DATA PROCESSING UNITS. https://www.nvidia.com/en-us/networking/products/data-processing-unit/.
[121] Martin Odersky, Lex Spoon, and Bill Venners. 2008. *Programming in Scala*. Artima Inc.

[122] P4.org. 2020. P4-16 Language Specification. https://p4.org/p4-spec/docs/P4-16-v1.2.1.pdf.

[123] Junghun Park, Hsiao-Rong Tyan, and C-C Jay Kuo. 2006. Internet Traffic Classification for Scalable QoS Provision. In *IEEE ICME '06*.

[124] Roberto Perdisci, Davide Ariu, Prahlad Fogla, Giorgio Giacinto, and Wenke Lee. 2009. McPAD: A Multiple Classifier System for Accurate Payload-Based Anomaly Detection. *Computer Networks* 53, 6 (2009), 864–881. https://doi.org/10.1016/j.comnet.2008.11.011

[125] Dan RK Ports and Jacob Nelson. 2019. When Should The Network Be The Computer?. In *ACM HotOS '19*. https://doi.org/10.1145/3317550.3321439

[126] Pascal Poupart, Zhitang Chen, Priyank Jaini, Fred Fung, Hengky Susanto, Yanhui Geng, Li Chen, Kai Chen, and Hao Jin. 2016. Online Flow Size Prediction for Improved Network Routing. In *IEEE ICNP '16*.

[127] Raghu Prabhakar, Yaqi Zhang, David Koeplinger, Matt Feldman, Tian Zhao, Stefan Hadjis, Ardavan Pedram, Christos Kozyrakis, and Kunle Olukotun. 2017. Plasticine: A Reconfigurable Architecture for Parallel Patterns. In *ACM/IEEE ISCA '17*. https://doi.org/10.1145/3079856.3080256

[128] Raghu Prabhakar, Yaqi Zhang, and Kunle Olukotun. 2020. Coarse-Grained Reconfigurable Architectures. In *NANO-CHIPS 2030*. Springer, 227–246. https://doi.org/10.1007/978-3-030-18338-7_14

[129] IO Visor Project. [n.d.]. XDP: eXpress Data Path. https://www.iovisor.org/technology/xdp. Accessed on 08/12/2021.

[130] Jack W Rae, Sergey Bartunov, and Timothy P Lillicrap. 2019. Meta-Learning Neural Bloom Filters. *arXiv:1906.04304* (2019).

[131] Alon Rashelbach, Ori Rottenstreich, and Mark Silberstein. 2020. A Computational Approach to Packet Classification. In *ACM SIGCOMM '20*. https://doi.org/10.1145/3387514.3405886

[132] Luigi Rizzo. 2012. Netmap: A Novel Framework for Fast Packet I/O. In *USENIX ATC '12*.

[133] Joshua Rosen, Neoklis Polyzotis, Vinayak Borkar, Yingyi Bu, Michael J Carey, Markus Weimer, Tyson Condie, and Raghu Ramakrishnan. 2013. Iterative mapreduce for large scale machine learning. *arXiv preprint arXiv:1303.3517* (2013).

[134] Alexander Rucker, Tushar Swamy, Muhammad Shahbaz, and Kunle Olukotun. 2019. Elastic RSS: Co-Scheduling Packets and Cores Using Programmable NICs. In *ACM APNet '19*. https://doi.org/10.1145/3343180.3343184

[135] Alexander Rucker, Matthew Vilim, Tian Zhao, Yaqi Zhang, Raghu Prabhakar, and Kunle Olukotun. 2021. *Capstan: A Vector RDA for Sparsity*. Association for Computing Machinery, New York, NY, USA, 1022–1035.

[136] Davide Sanvito, Giuseppe Siracusano, and Roberto Bifulco. 2018. Can the Network Be the AI Accelerator?. In *NetCompute '18*. https://doi.org/10.1145/3229591.3229594

[137] Amedeo Sapio, Marco Canini, Chen-Yu Ho, Jacob Nelson, Panos Kalnis, Changhoon Kim, Arvind Krishnamurthy, Masoud Moshref, Dan RK Ports, and Peter Richtárik. 2019. Scaling Distributed Machine Learning with In-Network Aggregation. *arXiv:1903.06701* (2019).

[138] Dipanjan Sarkar. 2018. Continuous Numeric Data – Strategies for Working with Continuous, Numerical Data. https://towardsdatascience.com/understanding-feature-engineering-part-1-continuous-numeric-data-da4e47099a7b. Accessed on 08/12/2021.

[139] Danfeng Shan, Fengyuan Ren, Peng Cheng, Ran Shu, and Chuanxiong Guo. 2018. Micro-burst in Data Centers: Observations, Analysis, and Mitigations. In *IEEE ICNP '18*. https://doi.org/10.1109/ICNP.2018.00019

[140] Ahmad Shawahna, Sadiq M Sait, and Aiman El-Maleh. 2018. Fpga-Based Accelerators of Deep Learning Networks for Learning and Classification: A Review. *IEEE Access* 7 (2018), 7823–7859.

[141] Justine Sherry, Sylvia Ratnasamy, and Justine Sherry At. 2012. A survey of enterprise middlebox deployments. (2012).

[142] Anton Shilov. [n.d.]. TSMC & Broadcom Develop 1700-mm2 CoWoS Interposer: 2x Larger Than Reticles. https://www.anandtech.com/show/15582/tsmc-broadcom-develop-1700-mm2-cowos-interposer-2x-larger-than-reticles. Accessed on 08/12/2021.

[143] Hartej Singh, Ming-Hau Lee, Guangming Lu, Fadi J Kurdahi, Nader Bagherzadeh, and Eliseu M Chaves Filho. 2000. MorphoSys: An Integrated Reconfigurable System for Data-Parallel and Computation-Intensive Applications. *IEEE Transactions on Computers '00* 49, 5 (2000), 465–481.

[144] Giuseppe Siracusano and Roberto Bifulco. 2018. In-Network Neural Networks. *arXiv:1801.05731* (2018).

[145] Arunan Sivanathan, Hassan Habibi Gharakheili, Franco Loi, Adam Radford, Chamith Wijenayake, Arun Vishwanath, and Vijay Sivaraman. 2018. Classifying IoT devices in Smart Environments Using Network Traffic characteristics. *IEEE Transactions on Mobile Computing '18* 18, 8 (2018), 1745–1759. https://doi.org/10.1109/TMC.2018.2866249

[146] Anirudh Sivaraman, Alvin Cheung, Mihai Budiu, Changhoon Kim, Mohammad Alizadeh, Hari Balakrishnan, George Varghese, Nick McKeown, and Steve Licking. 2016. Packet Transactions: High-Level Programming for Line-Rate Switches. In *ACM SIGCOMM '16*. https://doi.org/10.1145/2934872.2934900

[147] Anirudh Sivaraman, Suvinay Subramanian, Mohammad Alizadeh, Sharad Chole, Shang-Tse Chuang, Anurag Agrawal, Hari Balakrishnan, Tom Edsall, Sachin Katti, and Nick McKeown. 2016. Programmable Packet Scheduling at Line Rate. In *ACM SIGCOMM '16*. https://doi.org/10.1145/2934872.2934899

[148] Vibhaalakshmi Sivaraman, Srinivas Narayana, Ori Rottenstreich, Shan Muthukrishnan, and Jennifer Rexford. 2017. Heavy-Hitter Detection Entirely in the Data Plane. In *ACM SOSR '17*. https://doi.org/10.1145/3050220.3063772

[149] Mario Srouji, Jian Zhang, and Ruslan Salakhutdinov. 2018. Structured control nets for deep reinforcement learning. In *ICML '18*. PMLR, 4742–4751.

[150] Arvind K Sujeeth, HyoukJoong Lee, Kevin J Brown, Tiark Rompf, Hassan Chafi, Michael Wu, Anand R Atreya, Martin Odersky, and Kunle Olukotun. 2011. OptiML: an implicitly parallel domain-specific language for machine learning. In *ICML '11*.

[151] Jinsheng Sun and Moshe Zukerman. 2007. An Adaptive Neuron AQM for a Stable Internet. In *International Conference on Research in Networking '07*. Springer.

[152] Runyuan Sun, Bo Yang, Lizhi Peng, Zhenxiang Chen, Lei Zhang, and Shan Jing. 2010. Traffic Classification Using Probabilistic Neural Networks. In *IEEE ICNC '10*.

[153] Tuan A Tang, Lotfi Mhamdi, Des McLernon, Syed Ali Raza Zaidi, and Mounir Ghogho. 2016. Deep Learning Approach for Network Intrusion Detection in Software Defined Networking. In *IEEE WINCOM '16*.

[154] Mahbod Tavallaee, Ebrahim Bagheri, Wei Lu, and Ali A Ghorbani. 2009. A Detailed Analysis of the KDD CUP 99 Data Set. In *IEEE CISDA '09*.

[155] Simon Thompson. 2011. *Haskell: The Craft of Functional Programming*. Vol. 2. Addison-Wesley.

[156] Emanuel Todorov, Tom Erez, and Yuval Tassa. 2012. Mujoco: A Physics Engine for Model-Based Control. In *IEEE IROS '12*.

[157] C Van Rijsbergen. 1979. Information Retrieval: Theory and Practice. In *Proceedings of the Joint IBM/University of Newcastle upon Tyne Seminar on Data Base Systems '79*.

[158] Erwei Wang, James J Davis, Ruizhe Zhao, Ho-Cheung Ng, Xinyu Niu, Wayne Luk, Peter YK Cheung, and George A Constantinides. 2019. Deep Neural Network Approximation for Custom Hardware: Where We've Been, Where We're Going. *ACM Computing Surveys (CSUR) '19* 52, 2 (2019), 1–39. https://doi.org/10.1145/3309551

[159] Haining Wang, Danlu Zhang, and Kang G Shin. 2002. Detecting SYN Flooding Attacks. In *Twenty-First Annual Joint Conference of the IEEE Computer and Communications Societies '02*, Vol. 3. 1530–1539.

[160] Naigang Wang, Jungwook Choi, Daniel Brand, Chia-Yu Chen, and Kailash Gopalakrishnan. 2018. Training Deep Neural Networks with 8-bit Floating Point Numbers. In *NeurIPS '18*.

[161] Shuo Wang, Zhe Li, Caiwen Ding, Bo Yuan, Qinru Qiu, Yanzhi Wang, and Yun Liang. 2018. C-LSTM: Enabling Efficient LSTM using Structured Compression Techniques on FPGAs. In *ACM/SIGDA FPGA '18*. https://doi.org/10.1145/3174243.3174253

[162] Keith Winstein and Hari Balakrishnan. 2013. TCP Ex Machina: Computer-Generated Congestion Control. In *ACM SIGCOMM '13*. https://doi.org/10.1145/2534169.2486020

[163] Shihan Xiao, Haiyan Mao, Bo Wu, Wenjie Liu, and Fenglin Li. 2020. Neural Packet Routing. In *ACM NetAI '20*. https://doi.org/10.1145/3405671.3405813

[164] Xilinx. [n.d.]. Alveo U250 Data Center Accelerator Card. https://www.xilinx.com/products/boards-and-kits/alveo/u250.html. Accessed on 08/12/2021.

[165] Xilinx. [n.d.]. Xilinx: AXI Reference Guide. https://www.xilinx.com/support/documentation/ip_documentation/ug761_axi_reference_guide.pdf. Accessed on 08/12/2021.

[166] Xilinx. [n.d.]. Xilinx OpenNIC Shell. https://github.com/Xilinx/open-nic-shell. Accessed on 09/30/2021.

[167] Xilinx. [n.d.]. Xilinx: UltraScale+ Integrated 100G Ethernet Subsystem. https://www.xilinx.com/products/intellectual-property/cmac_usplus.html. Accessed on 08/12/2021.

[168] Zhaoqi Xiong and Noa Zilberman. 2019. Do Switches Dream of Machine Learning? Toward In-Network Classification. In *ACM HotNets '19*. https://doi.org/10.1145/3365609.3365864

[169] Francis Y Yan, Jestin Ma, Greg D Hill, Deepti Raghavan, Riad S Wahby, Philip Levis, and Keith Winstein. 2018. Pantheon: The Training Ground for Internet Congestion-Control Research. In *USENIX ATC '18*.

[170] Liangcheng Yu, John Sonchack, and Vincent Liu. 2020. Mantis: Reactive Programmable Switches. In *ACM SIGCOMM '20*. https://doi.org/10.1145/3387514.3405870

[171] Yasir Zaki, Thomas Pötsch, Jay Chen, Lakshminarayanan Subramanian, and Carmelita Görg. 2015. Adaptive Congestion Control for Unpredictable Cellular Networks. In *ACM SIGCOMM '15*. https://doi.org/10.1145/2785956.2787498

[172] Sebastian Zander, Thuy Nguyen, and Grenville Armitage. 2005. Automated Traffic Classification and Application Identification Using Machine Learning. In *IEEE LCN '05*.

[173] Jun Zhang, Chao Chen, Yang Xiang, Wanlei Zhou, and Yong Xiang. 2012. Internet Traffic Classification by Aggregating Correlated Naïve Bayes Predictions. *IEEE Transactions on Information Forensics and Security '12* 8, 1 (2012), 5–15. https://doi.org/10.1109/TIFS.2012.2223675

[174] Jun Zhang, Xiao Chen, Yang Xiang, Wanlei Zhou, and Jie Wu. 2014. Robust Network Traffic Classification. *IEEE/ACM Transactions on Networking '14* 23, 4 (2014), 1257–1270. https://doi.org/10.1109/TNET.2014.2320577

[175] Menghao Zhang, Guanyu Li, Shicheng Wang, Chang Liu, Ang Chen, Hongxin Hu, Guofei Gu, Qianqian Li, Mingwei Xu, and Jianping Wu. 2020. Poseidon: Mitigating Volumetric DDoS Attacks with Programmable Switches. In *NDSS '20*. https://doi.org/10.14722/ndss.2020.24007

[176] Yaqi Zhang, Alexander Rucker, Matthew Vilim, Raghu Prabhakar, William Hwang, and Kunle Olukotun. 2019. Scalable Interconnects for Reconfigurable Spatial Architectures. In *ACM/IEEE ISCA '19*. https://doi.org/10.1145/3307650.3322249

[177] Yaqi Zhang, Nathan Zhang, Tian Zhao, Matt Vilim, Muhammad Shahbaz, and Kunle Olukotun. 2021. SARA: Scaling a Reconfigurable Dataflow Accelerator. In *ACM/IEEE ISCA '21*. https://doi.org/10.1109/ISCA52012.2021.00085

[178] Hongtao Zhong, Kevin Fan, Scott Mahlke, and Michael Schlansker. 2005. A Distributed Control Path Architecture for VLIW Processors. In *IEEE PACT '05*.

[179] Chuan Zhou, Dongjie Di, Qingwei Chen, and Jian Guo. 2009. An Adaptive AQM Algorithm Based on Neuron Reinforcement Learning. In *IEEE ICCA '09*.